%% file: 0-main.tex
\documentclass[aps,prd,superscriptaddress,amsfonts,amssymb,amsmath,natbib,showpacs,nofootinbib,10pt]{revtex4-2}

\usepackage{xcolor}
\usepackage{comment}
\usepackage[normalem]{ulem}
\usepackage{graphicx}
\usepackage{dcolumn}
\usepackage{bm}
\usepackage{hyperref}
\usepackage[normalem]{ulem}
\hypersetup{
    colorlinks=true,        
    linkcolor=blue,         
    citecolor=blue,         
    urlcolor=blue           
}
\usepackage[mathlines]{lineno}
\usepackage{subfigure}
\usepackage{blindtext}
\usepackage{amsmath}
\usepackage{multirow}
\usepackage{lipsum}
\usepackage{enumitem}
\usepackage{booktabs}
\usepackage{float}
\usepackage{tabularx}
\usepackage{amssymb}
\usepackage{mathtools}
\usepackage{placeins}

\begin{document}

\title{Past and future isotropization in $f(R)$ dark energy models}

\author{Adnaan Nauthoo}
\email{nthadn001@myuct.ac.za}
\affiliation{Cosmology and Gravity Group, Department of Mathematics and Applied Mathematics, University of Cape Town, Rondebosch 7700, Cape Town, South Africa}

\author{Peter K. S. Dunsby}
\affiliation{Cosmology and Gravity Group, Department of Mathematics and Applied Mathematics, University of Cape Town, Rondebosch 7700, Cape Town, South Africa}
\affiliation{South African Astronomical Observatory, Observatory 7925, Cape Town, South Africa} 
\affiliation{Centre for Space Research, North-West University, Potchefstroom 2520, South Africa}
\author{Álvaro de la Cruz-Dombriz}
\affiliation{Cosmology and Gravity Group, Department of Mathematics and Applied Mathematics, University of Cape Town, Rondebosch 7700, Cape Town, South Africa}
\affiliation{Departamento de Física Fundamental, Universidad de Salamanca, 37008 Salamanca, Spain}

\date{\today}

\begin{abstract}
We present a dynamical system analysis for the homogeneous and anisotropic Bianchi I cosmology in the context of Hu-Sawicki $f(R)$ gravity with parameters $\{n,C_{1}\}=\{1,1\}$. Deriving the corresponding four-dimensional dynamical system from the field equations allows us to find the fixed points of the system in both the vacuum and matter cases alongside their stability. This approach also provides analytical expressions for both the cosmic expansion and shear of each fixed point. More importantly, it allows us to address one of the least satisfying aspects of the $\Lambda$CDM picture: the apparent fine-tuning of the initial state. We show that, for the chosen parameters, Hu-Sawicki gravity allows for  isotropization both in the past and in the future for both vacuum and matter, demonstrating that Hu-Sawicki gravity  may alleviate the anisotropy-related initial-condition problem for a subsequent inflationary phase. For the matter case, the analysis was subdivided into the dust and radiation cases, and we show that both cases exhibit past and future isotropization. The obtained results also revealed interesting aspects of the phase space such as the existence and phenomenology of lines and two-dimensional sheets of fixed points.
\end{abstract}

\maketitle

\input{01-Introduction}

\input{02-f_R_}
\input{02a-Field_equations}
\input{02b-Hu-Sawicki_Gravity}

\input{03-Dynamical_system}

\input{04-Results}
\input{04a-Vaccum}
\input{05-Matter}

\input{06-Conclusion}

\input{07-Acknowledgements}

\input{08-Appendix}

\bibliographystyle{unsrt}
\bibliography{refs}

\end{document}

%% file: 01-Introduction.tex
\section{Introduction}
Despite being formulated more than a hundred years ago, Einstein's General Theory of Relativity (GR) remains our best theory of gravity, surviving multiple observational tests ranging from solar system to galactic scales \cite{Ostriker:1995rn,*SupernovaSearchTeam:2003cyd,*SupernovaCosmologyProject:2003dcn,*SupernovaSearchTeam:2004lze,*SNLS:2005qlf,*WMAP:2003elm,*WMAP:2006bqn,*SDSS:2003eyi,*SDSS:2004kqt,*2dFGRS:2005yhx,*SDSS:2005xqv,*Blake:2005jd,*Jain:2003tba}.
The Friedmann-Lemaître-Robertson-Walker (FLRW)  solution, grounded in the Cosmological Principle — the postulate that the Universe is both homogeneous and isotropic on sufficiently large scales — has long served as the standard framework for modern cosmology. Nevertheless, increasingly precise observational data reveal that this description is not complete. Measurements from Type Ia supernovae, cosmic microwave background (CMB) anisotropies, and large-scale structure surveys consistently point to the necessity of additional physical ingredients to achieve concordance with observations. In particular, the phenomenon of late-time cosmic acceleration cannot be accommodated within the FLRW framework without the inclusion of an exotic energy component dubbed Dark Energy, with the simplest realization given by a cosmological constant ($\Lambda$) — which effectively introduces a repulsive contribution driving the accelerated expansion of the Universe \cite{SupernovaSearchTeam:1998fmf,SupernovaCosmologyProject:1998vns}. On the other hand, observations \cite{RubinFord1970, Planck2018} also indicate that a type of exotic matter must be present in the Universe in the form of some non-luminous component that acts gravitationally — Dark Matter — which, if non-relativistic (as it seems by many different observations), is usually referred to as Cold Dark Matter (CDM). Together, Dark Energy and Dark Matter are required ingredients in the $\Lambda$CDM (or Concordance) model, which remains the standard cosmological model due to its excellent agreement with current observational data. Despite its success, there are still issues with the $\Lambda$CDM model, such as the \textit{Hubble tension} \cite{DiValentino:2021izs}, where the inferred value of the Hubble constant from early-universe observations (e.g., CMB measurements) differs significantly from that obtained via late-universe measurements (e.g., local distance ladder techniques); and the \textit{cosmological constant problem}, where the inferred value of $\Lambda$ drastically differs with the theoretical estimate of the vacuum energy from Quantum Field Theory \cite{Weinberg1989}. These issues within the $\Lambda$CDM model and GR have created interest in modified gravity theories, which do not need the introduction of any ad-hoc dark components.

A paradigmatic modification of GR is  the so-called $f(R)$ theory — a family of scalar-tensor theories of gravity in which the GR gravitational action is generalized to a non-linear function of the Ricci scalar $R$ \cite{Sotiriou:2008rp, Starobinsky1980, DeFelice:2010aj}. Such corrections allow for richer and more complex cosmological dynamics that could potentially describe gravity without Dark Matter and Dark Energy \cite{Carloni:2006mr, PhysRevD.93.084016}. Several $f(R)$ theories have had success in describing the late-time acceleration of the Universe without the need for a cosmological constant or Dark Energy, including, among others, the so-called Starobinsky, Hu-Sawicki, and exponential models \cite{Starobinsky1980,Hu:2007nk,Nojiri:2006gh}. Due to the increased complexity of the resulting field equations, the dynamical systems approach \cite{Wainwright:1997, Carloni:2007br,Carloni:2006mr,Chakraborty:2021mcf} is often the most suitable and powerful way of studying such classes of models. In this work, we focus on the Hu-Sawicki model using the dynamical systems approach. Previous works have investigated the homogeneous and isotropic FLRW cosmology in the context of Hu-Sawicki gravity \cite{MacDevette:2022hts,PhysRevD.93.084016}. In this communication, we extend  such analyses to the homogeneous although anisotropic Bianchi I spacetimes \cite{bianchi1898,Misner:1969hg,ryan1975homogeneous}.

Anisotropic cosmological models are particularly useful in investigating early-universe anisotropies, as well as for studying shear dynamics and the eventual emergence of late-time anisotropies. In the context of $f(R)$ theories, anisotropic models also exhibit interesting features \cite{Leach:2006br}. One such feature is the possibility of isotropization in both the past and in the future, which occurs if the initial cosmological singularity is an isotropic past attractor. In standard GR, the Cosmic No-Hair Theorem \cite{PhysRevD.28.2118} guarantees that inflation drives the universe towards isotropy in the future by exponentially suppressing anisotropies. However, GR does not allow for past isotropization as the shear scalar $\sigma^{2}$ scales as $\sigma^{2} \propto a^{-6}$ ($a$ being the scale factor) \cite{Wainwright:1997}. This implies that $\sigma \to \infty$ as $a\to0$, preventing past isotropization. Consequently, an inflationary phase beginning from a Bianchi~I state
requires the initial shear to be sufficiently small. The presence of an
isotropic past attractor in an $f(R)$ theory is therefore potentially relevant to the initial-condition problem of inflation
because it provides a dynamical mechanism by which an anisotropic pre-inflationary
spacetime can approach isotropy. We stress, however, that past isotropization and
the generation of inflation are distinct questions. In the present work, we
investigate only the former: no inflationary sector is specified, and our
results should therefore be interpreted as identifying a possible
pre-inflationary isotropization mechanism rather than as providing a
complete model of inflation.

Brane-world cosmologies were amongst the first models which admitted an isotropic past attractor in the form of a simple FLRW solution. This was a significant breakthrough, as such an attractor emerged naturally, without the need of fine-tuning the initial conditions \cite{PhysRevD.63.104012,PhysRevD.64.063506,Dunsby:2003sr}. As $f(R)$ gravity emerged as a more promising theory, focus shifted on the latter. Since then, several works have investigated isotropization in the context of $f(R)$ gravity. For instance, the authors in \cite{Bhattacharya:2017cbn} showed that a Starobinsky-like inflation leads to future isotropization in Bianchi I models, while in \cite{Muller:2017nxg}, the authors demonstrated that $f(R) = R+R^{2}$ gravity admits isotropic past attractors. The case of $f(R) = R^{n}$ in Bianchi I was also studied in \cite{Leach:2006br} and the authors found that such models also admitted isotropic past attractors. Nevertheless, the behavior of anisotropies in $f(R)$ models like Hu-Sawicki  gravity have not yet been studied, particularly in a compactified dynamical framework. In this paper, we focus on the latter problem.

This paper is structured as follows: In Sec. \ref{sec:f(R)}, we briefly review $f(R)$ theories in the metric formalism before providing the relevant field equations in Sec. \ref{sec:f(R) Field Equations} for Bianchi I spacetimes. Then, in Sec. \ref{sec:Hu-Sawicki}, we provide a review of Hu-Sawicki gravity. We then specialize the Hu-Sawicki model for the parameters $\{n,C_{1}\}=\{1,1\}$ in Sec. \ref{sec:DS In BI with HS} and provide the relevant dynamical system equations. Subsequently, in Sec. \ref{sec:Results}, we tabulate our results by providing the fixed points, stability, scale factor and  shear of our dynamical system for both the vacuum and matter cases. Throughout the paper, we use metric signature $(-,+,+,+)$  and set $\kappa=8\pi G=c=1$ (unless otherwise stated). To streamline the notation, we will also sometimes abbreviate $f(R)$ to $f$.

%% file: 02-f_R_.tex
\section{$f(R)$ Theories}
\label{sec:f(R)}
The total action ${\cal S}$ in a generic $f(R)$ theory is given \cite{Sotiriou:2008rp,DeFelice:2010aj} by 
\begin{equation}
{\cal S}=\frac{1}{2}\int {\rm d}^4x \sqrt{- g} \, f(R)+\int {\cal L}_{M} {\rm d}^4x,
\label{action:f(R)}
\end{equation}
where $g$ is the determinant of the metric tensor $g_{\mu \nu}$  and ${\cal L}_M$ is the Lagrangian of the matter fields. By varying the action in (\ref{action:f(R)}) with respect to $g_{\mu \nu}$, one obtains the well-known equations of motion: 
\begin{equation}
f'(R) R_{\mu\nu} - \frac{1}{2} f(R) g_{\mu\nu} + \left( g_{\mu\nu} \Box - \nabla_\mu \nabla_\nu \right) f'(R) = T_{\mu\nu}^{M},
\label{EOM}
\end{equation}

\noindent
where $f'(R)$ denotes $\frac{{\rm d}f}{{\rm d}R}$, $\nabla_\mu$ is the usual covariant derivative, $\Box \equiv \nabla^\mu \nabla_\mu = g^{\mu\nu} \nabla_\mu \nabla_\nu $ and $T_{\mu\nu}^{M}$ is the usual matter energy-momentum tensor. Notice that (\ref{EOM}) reduces to GR for the case $f(R)=R$, since $f'(R)=1$. It is also possible to rewrite (\ref{EOM}) \textit{à la} Einstein,
\begin{equation}
    R_{\mu\nu} -\frac{1}{2}g_{\mu\nu}R = T_{\mu\nu}^{tot}, 
    \label{Field Equation}
\end{equation}
\noindent
by defining
\begin{equation}
T_{\mu\nu}^{tot} \equiv \frac{1}{f'}\left(T_{\mu\nu}^{M} + T_{\mu\nu}^{eff}\right), 
\end{equation}
where 
\begin{equation}
    T_{\mu\nu}^{eff} \equiv \frac{f - Rf'}{2} g_{\mu\nu} + \nabla _{\mu}\nabla _{\nu}f'-g_{\mu\nu}\Box f'
\end{equation}
is an effective energy-momentum tensor due to the modification of the gravitational action. The trace of (\ref{EOM}) then becomes
\begin{equation}
    f'R-2f+3\Box f'= T^M, 
    \label{Trace}
\end{equation}
where $T^M \equiv g^{\mu\nu}T_{\mu\nu}^{M}$. In this work, we will assume matter behaves like a barotropic perfect fluid, with pressure $p=w\rho$, where $\rho$ represents the energy density and $w$ the constant equation of state parameter. Consequently, we have
\begin{equation}
    T_{\mu\nu}^{M} = (\rho + p)u_{\mu}u_{\nu} + pg_{\mu \nu} \, , 
    \label{Perfect Fluid}
\end{equation}
where $u^{\mu}$ is the fluid's four-velocity and is normalized such that $u^{\mu}u_{\mu} =-1$. Energy-momentum conservation implies that $\nabla^{\mu}T_{\mu\nu}^{M}=0$. Hence, the vanishing divergence of (\ref{Perfect Fluid}) leads to conservation of energy of our system.

%% file: 02a-Field_equations.tex
\subsection{$f(R)$ field equations in Bianchi I Cosmology}
\label{sec:f(R) Field Equations}
The four-dimensional metric for a Bianchi I spacetime is given by 
\begin{equation}
    {\rm d}s^{2}= -{\rm d}t^{2}+a_{1}^{2}(t){\rm d}x^{2}+a_{2}^{2}(t){\rm d}y^{2}+a_{3}^{2}(t){\rm d}z^{2},
\end{equation}
with $a_{1,2,3}(t)$ 
being the scale factors in the $x$, $y$ and $z$ directions respectively. It is convenient to introduce the average scale factor $a(t)=(a_{1}a_{2}a_{3})^{\frac{1}{3}}$. In the Locally Rotationally Symmetric (LRS) subclass, two of the
directional scale factors coincide, leaving a single independent anisotropy degree of freedom. In what follows, we will restrict ourselves to the LRS subclass. This choice is motivated by the fact that a single independent anisotropy component yields a system that is
analytically and numerically tractable. From the $f(R)$ field equations \eqref{EOM}, one can then derive the equations that govern the dynamics of a LRS Bianchi I cosmology. These equations are given by~\cite{Chakraborty:2018ost}:

\begin{enumerate}[label=\textit{\arabic*.}]
    \item \textit{Friedmann equation}:
    \begin{equation}
    3H^2=\frac{1}{f^{\prime}}\left(\rho+\frac{Rf^{\prime}-f}{2}-3Hf^{\prime\prime}\dot{R}\right)+ {\sigma^2},
    \label{Friedmann}
    \end{equation}

    \item \textit{Raychaudhuri equation}:
    \begin{equation}
        2\dot{H}+3H^2 = -\frac{1}{f^{\prime}}\left[w\rho+\dot{R}^2f^{\prime\prime\prime}+\left(2H\dot{R} +\ddot{R}\right)f^{\prime\prime} -\frac{Rf^{\prime}-f}{2}\right] -\sigma^2,
    \label{Raychaudauri}
    \end{equation}

    \item \textit{Energy conservation equation}:
    \begin{equation}
    \dot{\rho} +  3H\left(1+w\right ) \rho =0,
    \label{Conservation}
    \end{equation}

    \item \textit{Gauss-Codazzi equation}:
    \begin{equation}   \dot{\sigma}+\left(3H+\frac{\dot{R}f''}{f'}\right){\sigma}=0,
    \label{Gauss-Codazzi}
    \end{equation}
\end{enumerate}
where as usual, $H = \frac{\dot{a}(t)}{a(t)} $ denotes the average Hubble parameter. Note that a dot derivative represents the derivative with respect to cosmic time $t$ and that $\sigma$ is defined via the symmetric shear tensor $\sigma_{\mu \nu}$, as $\sigma^{2} = \frac{1}{2}\sigma_{\mu \nu}\sigma^{\mu\nu}$, where $\sigma_{\mu\nu}=h_{\mu}{}^{\alpha}h_{\nu}{}^{\beta}\nabla_{(\alpha}u_{\beta)}-\frac{1}{3}\theta\,h_{\mu\nu}$, with $h_{\mu\nu}=g_{\mu\nu}+u_\mu u_\nu$ the spatial projector and $\theta=\nabla_{\alpha}u^{\alpha}$ the volume expansion scalar.

In addition to equations (\ref{Friedmann})–(\ref{Gauss-Codazzi}), one needs one additional equation that governs the dynamical evolution of $R$ to close this system. The latter role will be played by (\ref{Trace}). As such, equations (\ref{Friedmann})–(\ref{Gauss-Codazzi}), together with equation (\ref{Trace}), form a closed system that can be recast into an autonomous system of first-order ordinary differential equations (ODEs) as will be seen in Sec. \ref{sec:DS In BI with HS} .     

%% file: 02b-Hu-Sawicki_Gravity.tex
\subsection{Hu-Sawicki gravity}
\label{sec:Hu-Sawicki}
The Hu-Sawicki (henceforth HS) model \cite{Hu:2007nk} is considered to be one of the most promising classes of $f(R)$ models; in what follows we will focus on the HS model. This choice is physically motivated, as unlike many toy $f(R)$ models, the HS construction is designed to: explain late-time acceleration without requiring a cosmological constant, reproduce the $\Lambda$CDM expansion history in the future, satisfy local and cosmological tests, and avoid the instabilities that plague many $f(R)$ models \cite{Sawicki:2007tf,Amendola:2006we,Amendola:2007nt}. Although other viable $f(R)$ models --- for e.g.,\ the Starobinsky (Dark Energy) and Appleby--Battye constructions \cite{Starobinsky:2007hu,Appleby:2007vb} --- also satisfy these conditions, the HS model has become a standard benchmark in $f(R)$ cosmology and is among the most widely used parametrizations in phenomenological and numerical studies. We therefore adopt it here as a representative member of the viable class and analyse the asymptotic behaviour of anisotropies, with particular emphasis on whether isotropy can be dynamically achieved in the early Universe. 

In general, for an $f(R)$ theory to be viable, the following criteria need to be met: 
\begin{itemize}
    \item $f(R) \to R$ for $R \to 0$ so as to recover GR in low-curvature regimes.
    \item $f(R) \to R-2\Lambda$ ($\Lambda>0$) for $R \to \infty$ so as to recover GR in high-curvature regimes.
    \item $f'(R)>0$ for all $R$ so that gravity remains solely attractive.
    \item $f''(R)>0$ for all $R$ so that at no point in time there is an unbounded growth of curvature perturbations, effectively avoiding the so-called Dolgov-Kawasaki instability \cite{Dolgov:2003px}.
\end{itemize}

In its most generic form, the HS model is a three-parameter theory $\{n,C_{1},C_{2}\}$ of the form
\begin{equation}
\label{HS f(R)}
    f(R) = R - \frac{{C_{1}}H_{0}^{2} (R/H_{0}^{2})^{n}}{{C_{2}} (R/H_{0}^{2})^{n} + 1},
\end{equation}
with $H_{0}$ being the Hubble parameter evaluated today.
Notice that in the limit $R \to \infty$, we have $\Lambda = \frac{H_{0}^{2}C_{1}}{2 C_{2}}$ from the criterion we had above, implying that $\frac{C_{1}}{C_{2}}$ has to be positive. We will exclusively look at the HS model for $n=C_{1} = 1$ in the following as it is the only model whose cosmic expansion equations can be turned into an autonomous dynamical system, as was pointed out in \cite{Kandhai:2015pyr}. Choosing to work for the particular case of $n=1$ does not make our subsequent analysis any less relevant, since the statistical significance of $n=1$ is comparable to other values of $n>1$ when compared to data~\cite{PhysRevD.93.084016,Santos,Perez-Romero:2017njc}.

It is important to distinguish the dimensionless
phase-space structure from the physical curvature scale associated with
a given trajectory. For $n=C_1=1$, the relevant dimensionless quantity is
\begin{equation}
 \chi=C_2R/H_0^2.
\end{equation}
The limits $\chi\ll1$ and $\chi\gg1$ correspond to the two GR-like
asymptotic regimes of the Hu--Sawicki function, while deviations from
these limits occur around $\chi=O(1)$. As will be shown in Sec.~\ref{sec:DS In BI with HS}, once the normalized variables and the closure relation are introduced, $C_2$ no longer appears explicitly in the autonomous phase-space equations. Nevertheless, $C_2$ fixes the conversion between the dimensionless curvature and the corresponding physical curvature through 
\begin{equation}
R=\frac{H_0^2}{C_2}\chi.
\end{equation}
The phase-space analysis therefore determines the curvature relative to the Hu--Sawicki transition scale, while assigning an absolute physical curvature scale requires specifying $C_2$. The compact autonomous system determines whether
past-isotropizing trajectories exist, but does not by itself imply that
the standard late-time choice of the Hu--Sawicki parameters places this
regime at the curvature scale associated with a conventional
inflationary epoch. A complete pre-inflationary model would require the
physical curvature scale to be specified and matched to an inflationary
sector. Our result should therefore be understood as a statement about
the existence of past-isotropizing sectors in the Hu--Sawicki phase
space, rather than as a complete model connecting the late-time
Hu--Sawicki theory directly to inflation.

%% file: 03-Dynamical_system.tex
\section{Dynamical System in Bianchi I in Hu-Sawicki Gravity }
\label{sec:DS In BI with HS}

We now define a set of \emph{compact} normalized variables that allows us to recast (\ref{Friedmann}), (\ref{Raychaudauri}), (\ref{Conservation}) and (\ref{Gauss-Codazzi}) in a system of autonomous first-order ODEs. As previously discussed, to close the system, we additionally require (\ref{Trace}). We start by first recasting the Friedmann equation,  (\ref{Friedmann}), as
\begin{equation}
\label{Compact Friedmann}
 D^2 \equiv  \left(3H + \frac{3}{2}\frac{\dot{f'}}{f'}\right)^{2} + \frac{3}{2}\left(\frac{f}{f'}\right) = \frac{3\rho}{f'} + \frac{3}{2}R + \frac{9}{4}\left(\frac{\dot{f'}}{f'}\right)^2 + 3\sigma^{2},
\end{equation}
which defines the normalisation variable $D$. In the physical sector, $R>0$, $\rho\geq 0$ and
$\sigma^{2}\geq 0$, which implies $D^{2}>0$ (with $f'> 0$), so that $D$ is real.
We therefore choose the positive branch, $D>0$, which fixes the orientation of the phase-space time
introduced below.

Next, we divide both sides by $D^2$, which naturally leads us to define a set of compact dynamical variables 
\begin{equation}
\label{compact variables}
 x=\frac{3}{2}\frac{\dot{f}^{\prime}}{f^{\prime}}\frac{1}{D}, \qquad v=\frac{3}{2}\frac{R}{D^{2}}, \qquad y=\frac{3}{2}\frac{f}{f^{\prime}}\frac{1}{D^{2}}, \qquad
 \Omega=\frac{3\rho}{f^{\prime}}\frac{1}{D^{2}}, \qquad Q=\frac{3H}{D}, \qquad \Sigma=\frac{3\sigma^{2}}{D^{2}},
\end{equation}
which are now related via (\ref{Compact Friedmann}) by two constrained equations
\begin{subequations}
\begin{eqnarray}
&& (Q+x)^{2}+y=1 \label{Friedmann constraint 1},\\
&& \Omega+v+x^{2}+\Sigma=1 \label{Friedmann constraint 2}.
\end{eqnarray}
\end{subequations}
From the definitions in (\ref{compact variables}) and from the two constraints (\ref{Friedmann constraint 1}) and (\ref{Friedmann constraint 2}), we conclude that the dynamical variables \eqref{compact variables} are compact and bounded on the following finite domains
\begin{equation}
\begin{aligned}
-1 \leq x \leq 1, \quad
   0 \leq \Omega \leq 1, \quad
  -2 \leq Q \leq 2, 
\quad
0 \leq v \leq 1, \quad
  0 \leq y \leq 1, \quad
  0 \leq \Sigma \leq 1.
  \label{Compact Variables Range}
\end{aligned}
\end{equation}

To write an autonomous dynamical system, we now  define a new phase-space time variable $\tau$ given by
\begin{equation}
    {\rm d}\tau \equiv D\, {\rm d}t,
\end{equation}
which, since $D>0$, is a monotonically increasing function of $t$.
Differentiating the dynamical variables in (\ref{compact variables}) with respect to $\tau$ and using (\ref{Friedmann})–(\ref{Gauss-Codazzi}) together with (\ref{Trace}) results in a system of six first-order autonomous differential equations. Due to the two constraints (\ref{Friedmann constraint 1}) and (\ref{Friedmann constraint 2}), two of the dynamical variables become redundant and we can therefore reduce the dimension of the dynamical variables to only four, effectively collapsing the 6D system to a 4D one. Here, we choose to eliminate $y$ and $\Omega$ and write the dynamical system as follows:

\begin{widetext}
\begin{subequations}\label{Compact DS}
\begin{eqnarray}
       && \frac{{\rm d}v}{{\rm d}\tau} = -\frac{v}{3}\Bigg\{(Q+x)\left[2v - (1+3w)(1-x^{2}-v) + 4xQ\right] - 2Q - 4x + 2x\Gamma(v-1) + 3(-1+w)(Q+x)\Sigma \Bigg\},
       \\
       && \frac{{\rm d}x}{{\rm d}\tau} = \frac{1}{6}\Bigg\{-2x^{2}v\Gamma + (1-3w)(1-x^{2}-v) + 2v + 4\left(x^{2}-1\right)\left(1-Q^{2}-xQ\right) + x(Q+x)\left[(1+3w)(1-x^{2}-v) - 2v\right]\nonumber\\
       && \qquad +\left[-1 +3w + 3x(Q+x)(1-w)\right]\Sigma \Bigg\},
       \\
       && \frac{{\rm d}Q}{{\rm d}\tau} = \frac{1}{6}\Bigg\{-4xQ^{3} + xQ(5+3w)(1-xQ) - Q^{2}(1-3w) - Qx^{3}(1+3w) - 3vQ(1+w)(Q+x)\nonumber\\
       && \qquad + 2v(1-Qx\Gamma) - \left[2+3Q(-1+w)(Q+x)\right]\Sigma\Bigg\},
       \\
       && \frac{{\rm d}\Sigma}{{\rm d}\tau} =
       -\frac{\Sigma}{3} \Bigg\{ 4Q^{2}x+x\left[(1+3w)(-1+x^{2}) + v(3+3w) + 3(-1+w)\Sigma\right] \nonumber \\
       && \qquad +Q\left[3+3v(1+w)+5x^{2}- 3\Sigma +3w(-1+x^{2}+\Sigma) \right] 
       - 2\Gamma v x \Bigg\},
    \end{eqnarray}
\end{subequations}
\end{widetext}
where we define an auxiliary quantity
\begin{equation}
    \Gamma \equiv \frac{f'}{Rf''}\,,
\end{equation}
which is explicitly dependent on the functional form of $f(R)$. $\Gamma$ essentially dictates the closure of our dynamical system. For any closed system, $\Gamma$ must be expressible as a function of the dynamical variables. With this in mind, notice that
\begin{equation}\label{Invert relation}
    \frac{v}{y} = \frac{Rf'}{f}.
\end{equation}
As such, closure can be obtained by solving (\ref{Invert relation}) for $R=R\left(\frac{v}{y}\right)$ and using it to write $\Gamma=\Gamma\left(\frac{v}{y}\right)$, provided that the relation (\ref{Invert relation}) is invertible. For the HS model with parameters $n=C_{1}=1$ (leaving $C_2$ arbitrary, since
it drops out of the autonomous evolution equations once this choice is made), \eqref{Invert relation} is invertible and one finds that
\begin{eqnarray}
    \Gamma = \frac{1}{2}\frac{vy}{(v-y)^2}.
\end{eqnarray}
Nonetheless, with the above expression for $\Gamma$, the dynamical system in \eqref{Compact DS} becomes singular at $v=y$ on the 3D hypersurface given by 
\begin{eqnarray}\label{z}
    z\equiv(v-y)=v+(Q+x)^{2}-1=0.
\end{eqnarray}
To this end, we redefine the phase-space time variable as
\begin{eqnarray}\label{time_redef}
    {\rm d}\eta = \frac{{\rm d}\tau}{\left[v+(Q+x)^{2}-1\right]^2},
\end{eqnarray}
so that with respect to this redefined time variable $\eta$, the new dynamical system is regular everywhere. The explicit form of the dynamical system in the compact phase space in the presence of a generic perfect fluid under the HS model with $n=C_{1}=1$ is then as follows:  

\begin{widetext}
\begin{subequations}\label{Reg Compact DS}
\begin{eqnarray}
&& \frac{{\rm d}v}{{\rm d}\eta} = -\frac{1}{3}vz^{2}\Bigg\{(Q+x)\left[2v-(1+3w)(1-x^{2}-v)+4xQ\right]-2Q-4x + 3 (-1+w)(Q+x)\Sigma\Bigg\}\\
&& - \frac{1}{3}v^{2}x(v-z)(v-1), \nonumber
\\
&& \frac{{\rm d}x}{{\rm d}\eta} = \frac{1}{6}z^{2}\Bigg\{(1-3w)(1-x^{2}-v)+2v+4(x^{2}-1)(1-Q^{2}-xQ)+x(Q+x)\left[(1+3w)(1-x^{2}-v)-2v\right]\\
&& +\left[-1 +3w + 3x(Q+x)(1-w)\right]\Sigma \Bigg\} - \frac{1}{6}x^{2}v^{2}(v-z), \nonumber
\\
&& \frac{{\rm d}Q}{{\rm d}\eta} = \frac{1}{6}z^{2}\Bigg\{-4xQ^{3}+xQ(5+3w)(1-xQ)-Q^{2}(1-3w)-Qx^{3}(1+3w)-3vQ(1+w)(Q+x)\\
&& +2v -\left[2+3Q(-1+w)(Q+x)\right]\Sigma\Bigg\} - \frac{1}{6}v^{2}(v-z)Qx, \nonumber
\\
&& \frac{{\rm d}\Sigma}{{\rm d}\eta} = -\frac{1}{3}z^{2}\Sigma \Bigg\{
4Q^{2}x + x \left[ (1+3w)(-1+x^{2}) + 3v(1+w) + 3(-1+w)\Sigma \right] \\
&& + Q \left[ 3 + 3v(1+w) 
    + 5x^{2} - 3\Sigma + 3w(-1+x^{2}+\Sigma) \right]
\Bigg\} - {\frac{1}{3}}v^{2}x(v-z)\Sigma \nonumber.
\end{eqnarray}
\end{subequations}
\end{widetext}

%% file: 04-Results.tex
\section{Results}
\label{sec:Results}
In this section we present the results of the dynamical system \eqref{Reg Compact DS} by tabulating the fixed points, scale factor and shear. We consider two particular cases: the vacuum case ($\Omega = 0$) and the matter case, for which we shall further subdivide the analysis for dust ($w=0$) and radiation ($w=\frac{1}{3}$). Note that in the results that follow, many fixed points were discarded in both the vacuum and matter cases, as they did not satisfy the bounds of the dynamical variables given in (\ref{Compact Variables Range}).

To obtain the analytical form of $a(t)$ for each of the fixed points, we substitute the Friedmann equation,  (\ref{Friedmann}), into the Raychaudhuri equation, (\ref{Raychaudauri}), and use the definitions in (\ref{compact variables}) to obtain an integrable equation for $a(t)$. We simply state $a(t)$ here and provide Appendix \ref{Appendix A} for an in-depth derivation. The scale factor at each fixed point is given by 
\begin{equation}
    a(t) = a_{0} (t-t_{0})^{\frac{2}{\alpha}},
    \label{Scale Factor}
\end{equation}
where $a_0$ is a normalization constant, $t_{0}$ denotes some initial time (which we will take to be the Big Bang) and $\alpha$ defined as
\begin{equation}
    \left.\alpha \equiv \frac{1}{Q^{2}}[4\Omega-8xQ+2v-4y+6\Sigma ]\right |_{P_{0}},
    \label{Alpha Definition}
\end{equation}
with $\alpha >0$ and $P_{0}$ being the coordinates of an arbitrary fixed point.
Two cases need to be approached with caution: $\alpha = 0$, which results in an undefined expression in (\ref{Scale Factor}); and $Q=0$, which results in an undefined expression in (\ref{Alpha Definition}). For $Q=0$, we have $H=0$, which implies that the scale factor is constant, $a(t) = a_{0}$, an Einstein static solution. For $\alpha=0$, we have $\dot{H} =0$ (see (\ref{H main eq}) in Appendix \ref{Appendix A} for details), implying that the scale factor takes the form $a(t)=a_{0}{\rm e}^{H_{\rm dS}t}$, a de Sitter solution, with $H_{\rm dS}$ being the constant Hubble parameter in the de Sitter (dS) case.

The shear $\sigma(t)$ at each fixed point can be similarly obtained by integrating the Gauss-Codazzi equation, (\ref{Gauss-Codazzi}). We again state $\sigma(t)$ and provide Appendix \ref{Appendix B}
 for a full derivation. The  shear at each fixed point is given by
 \begin{equation}
     \sigma(t)= \sigma_{0}{a_{0}}^{-\gamma}(t-t_{0})^{-\frac{2\gamma}{\alpha}},
     \label{Shear with a(t)}
 \end{equation}
with $\gamma$ defined as 
\begin{equation}
    \gamma \equiv \left. 3+\frac{2x}{Q}\right |_{P_{0}}.
    \label{Gamma Definition}
\end{equation}
The cases of $Q=0$ and $\alpha=0$ must again be treated with caution. In such instances, instead of using \eqref{Scale Factor} to obtain \eqref{Shear with a(t)}, we use the respective forms of the scale factor (Einstein static for $Q=0$ and de Sitter for $\alpha=0$) to obtain an expression for the shear at these points (see Appendix \ref{Appendix B} for details). 
\subsection{Physical Interpretation of Fixed Points}
Before presenting the fixed points in Tables~\ref{Table Vacuum Case}--\ref{Table Radiation}, it is useful to summarize how the
compact variables encode the physical cosmology:
\begin{itemize}
    \item 
    
    \textit{Expansion versus contraction.}
    The variable $Q$ is proportional to the Hubble parameter, $Q = 3H/D$.
    Since $D$ is defined as a positive normalization, the sign of $Q$ coincides with the sign of $H$.
    Thus, $Q>0$ corresponds to an expanding phase, $Q<0$ to a contracting phase, and $Q=0$ to
    $H=0$ (an Einstein static or turning-point configuration).
    
    \item 

    \textit{Isotropy.}
    The variable $\Sigma$ is the  normalized shear, $\Sigma = 3\sigma^{2}/D^{2}$.
    Hence, $\Sigma=0$ is precisely the isotropic (FLRW) submanifold, while $\Sigma>0$ corresponds
    to anisotropic Bianchi~I states. In particular, a fixed point with $\Sigma=0$ is isotropic
    for all times, whereas a fixed point with $\Sigma>0$ is anisotropic, with the late/early-time
    behavior determined by the associated shear, $\sigma(t)$.

    \item
    
    \textit{Power-law, de Sitter, and static solutions.}
    At any fixed point $P_{0}$, the quantity $\alpha(P_{0})$ defined in \eqref{Alpha Definition} is constant. For $\alpha(P_{0})>0$
    with $Q\neq 0$, \eqref{Scale Factor} gives a power-law scale factor $a(t)\propto (t-t_{0})^{m}$ with exponent $m=2/\alpha(P_{0})$.
    The special case $Q=0$ implies $H=0$ and hence a static solution $a(t)=a_{0}$.
    When $\alpha(P_{0})=0$ (with $Q\neq 0$), the solution is a de Sitter spacetime, $a(t)\propto e^{H_{\mathrm{dS}} t}$.
    
\end{itemize}

In the expanding branch ($Q>0$), a de Sitter equilibrium corresponds to accelerated late-time expansion
($\ddot a>0$), and therefore provides the natural fixed-point representation of a Dark Energy-like regime
in the HS model.

%% file: 04a-Vaccum.tex
\subsection{Vacuum Case}
\label{sec:Vacuum}
The vacuum case is obtained by setting $\Omega = 0$. Hence, the constraint (\ref{Friedmann constraint 2}) simplifies to give
\begin{equation}
    v+x^{2}+\Sigma =1,
    \label{Vaccum Constraint}
\end{equation}
 allowing us to recast our dynamical system (\ref{Reg Compact DS}) in terms of only three dynamical variables. This effectively collapses the 4D system into a 3D one, making one of the equations in (\ref{Reg Compact DS}) redundant. The choice of expressing one variable in terms of the other two in (\ref{Vaccum Constraint}) dictates which equation in (\ref{Reg Compact DS}) can be effectively discarded. Here, we use (\ref{Vaccum Constraint}) to express $v$ in terms of $x$ and $\Sigma$, making the $\frac{{\rm d}v}{{\rm d}\eta}$ equation redundant. We then solve for the fixed points of the system by setting $\frac{{\rm d}x}{{\rm d}\eta}=\frac{{\rm d}Q}{{\rm d}\eta}=\frac{{\rm d}\Sigma}{{\rm d}\eta}=0$. These are shown in Table \ref{Table Vacuum Case}.
\begin{table}[h]
	\centering
    \resizebox{\textwidth}{!}{%
  
	\begin{tabular}{|c|c|c|c|c|}
    
		\hline
		Point & $(\Sigma,x,v,Q,\Omega, y)$ & Stability & Scale factor, $a(t)$ & Shear, $\sigma(t)$\\
		
		\hline
		$\mathcal{L}_{1\pm}$ & $\left( 1-x^{2},x,0,\pm 1 -x,0,0 \right)$ & Unstable/Stable &
    $\begin{cases}
    
    & \text{$a_{0}(t-t_{0})^{\tfrac{\pm1-x}{\mp3+x}}$}, \\
    & \text{$a_{0}$ for $x = \pm1$}\\ 
    
    \end{cases}$
    
    & 
    $\begin{cases}
    
    & \text{$ \sigma_{0}a_{0}^{{\tfrac{\pm3-x}{\mp1+x}}}(t-t_{0})^{-1}$}, \\
    & \text{$0$ for  $x = \pm1$}\\ 
    
    \end{cases}$
    \\
		
	    \hline

        $\mathcal{L}_{2}\,^{\dagger}$ & $\left( Q^{2},0,1-Q^{2},Q,0,1-Q^{2} \right)$ & 
        
    $\begin{cases}
    
    & \text{Saddle for $0\leq Q \lesssim 0.37$}, \\
    & \text{Unstable otherwise}\\
    
    \end{cases}$ &
    
    $\begin{cases}
    
    & \text{$a_{0}(t-t_{0})^{\tfrac{Q^{2}}{{4Q^{2}-1}}}$}, \\
    & \text{$a_{0}{\rm e}^{H_{\rm dS}t}$ for $Q = \pm\tfrac{1}{2}$},\\ 
    &\text{$a_{0}$ for $Q = 0$}

    \end{cases}$
    
    & 
    $\begin{cases}
    
    & \text{$\tfrac{\sigma_{0}}{{a_{0}^{3}}} {(t-t_{0})}^{\tfrac{3Q^{2}}{{1-4Q^{2}}}}$}, \\
    & \text{$\tfrac{\sigma_{0}}{{a_{0}}^{3}}{\rm e}^{-3H_{\rm dS}t}$ for $Q=\pm\tfrac{1}{2}$}, \\
    & \text{$0$ for $Q=0$}\\ 
    
    \end{cases}$
    \\

    \hline
    $\mathcal{A}_{\pm}$ & $\left(0,\pm 1,0,\pm \tfrac{1}{2},0,-\frac{5}{4} \right)$ & Saddle & $a_{0}\sqrt{t-t_{0}}$ & 0\\

    \hline
    $\mathcal{B}_{\pm}$ & $\left(0,\pm 1,0,0,0,0 \right)$ & Unstable/Stable & $a_{0}$ & 0\\

    \hline

    $\mathcal{C}_{\pm}$ & $\left(0,\mp 1,0,\pm 2,0,0 \right)$ & Unstable/Stable & $a_{0}\sqrt{t-t_{0}}$ & 0\\

    \hline

    $\mathcal{D}_{\pm}$ & $\left(0,\pm \frac{1}{3},\frac{8}{9},\pm \frac{2}{3},0,0 \right)$ & Stable/Unstable & $a_{0} {\rm e}^{H_{\rm dS}t}$ & 0 \\

    \hline

    $\mathcal{E}_{\pm}$ & $\left(1,0,0,\pm 1,0,0 \right)$ & Unstable/Stable & $a_{0}(t-t_{0})^{1/3}$ & $\tfrac{\sigma_{0}}{{a_{0}}^{3}(t-t_{0})}$  \\

    \hline

    $\mathcal{F}$ & $\left(0,0,1,0,0,1 \right)$ & Saddle & $a_{0}$ & $0$  \\

    \hline

    $\mathcal{G_{\pm}}$ & $\left(0,0,1,\pm \frac{1}{\sqrt{2}},0,\frac{1}{2} \right)$ & Stable/Unstable & $a_{0} {\rm e}^{H_{\rm dS}t}$ & $0$  \\

    \hline

	    \end{tabular}}
		\caption{Fixed points for the HS model with the parameters $\{n,C_{1}\} = \{1,1\}$ in vacuum ($\Omega = 0$). Notice that this system has three lines of fixed points, $\mathcal{L}_{1\pm}$ and $\mathcal{L}_{2}$ and thirteen fixed points. Where unspecified, the dynamical variables can take all the values on the interval over which they were defined in (\ref{Compact Variables Range}). $^{\dagger}$Here, for the line $\mathcal{L}_{2}$, the range for $Q$ is $|Q| \leq 1$ as $\Sigma = Q^{2}$ on the line and $\Sigma$ is bounded on the interval $[0,1]$, which puts a restriction on the possible values $Q$ can take.}
		\label{Table Vacuum Case}
\end{table}

As can be seen from Table \ref{Table Vacuum Case}, the vacuum case has three lines of fixed points $\mathcal{L}_{1 \pm}$ and $\mathcal{L}_{2}$ and thirteen fixed points $\mathcal{A_{\pm}},\mathcal{B_{\pm}},\mathcal{C_{\pm}},\mathcal{D_{\pm}},\mathcal{E_{\pm}}, \mathcal{F} \text{ and } \mathcal{G_{\pm}}$. Among these, $\mathcal{A_{\pm}}$, $\mathcal{D_{\pm}}$ and $\mathcal{G_{\pm}}$ are isolated fixed points. The cosmologies for all the fixed points are described by three scale factor solutions: a power-law of the form $a(t)\sim(t-t_{0})^{m}$ (for some $m \neq0$), a de Sitter universe of the form $a(t) \sim {\rm e}^{H_{\rm dS}t}$ and a static universe of the form $a(t)\sim a_{0}$. The stability of the fixed points was determined by the eigenvalues of the Jacobian matrix of the system. However, it is to be noted that this method can only be used when none of the eigenvalues of the Jacobian are 0. In our case, some of the fixed points did have vanishing eigenvalues, in which case other methods needed to be used to determine the stability. For an in-depth analysis of how the stability was determined, see Appendix \ref{Appendix C}. 

We should also expect to recover some of the fixed points that the authors in \cite{MacDevette:2022hts} found. Their analysis was performed on the isotropic (i.e., $\Sigma=0$) and homogeneous FLRW cosmology which, in principle, includes non-zero spatial curvature $K$. Since our analysis is focused on Bianchi I, for which $K=0$ (but $\Sigma$ not necessarily $0$), all fixed points that obey $\Sigma=K=0$ should appear in both the FLRW and Bianchi I cases. Indeed, two such pairs of fixed points were recovered: $\mathcal{D_{\pm}}$ and $\mathcal{G_{\pm}}$. For $\mathcal{D_{\pm}}$, we obtained the same stability and cosmic expansion as found in \cite{MacDevette:2022hts}, while for $\mathcal{G_{\pm}}$, we obtained the same cosmic expansion and were able to additionally deduce that $\mathcal{G_{\pm}}$ must be stable and unstable respectively.    

As for the shear, the lines $\mathcal{L}_{1\pm}$ admit a power-law and a static solution depending on the variable $x$, which parametrizes the lines. These lines are non-isotropic for all values of $x$ with the exception of $x=\pm1$ (the static case), which corresponds to the point where the lines cross the $\Sigma=0$ plane. Hence, for $x=\pm1$, the lines $\mathcal{L}_{1\pm}$ are isotropic for all times. The power-law case allows for late-time isotropization ($\sigma \to 0$ as $t\to\infty$) due to the $(t-t_{0})^{-1}$ factor which makes $\sigma(t)\to0$ as $t\to\infty$. For the line $\mathcal{L}_{2}$, it admits a power-law, de Sitter and static solutions. It is also isotropic for all times when $Q=0$ (the static case). For the power-law and de Sitter solutions, we see that these models can isotropize as $t\to\infty.$ In particular, for the power-law case solving $\frac{3Q^{2}}{1-4Q^{2}}<0$ constraint to $|Q| \leq 1$ (see the $^{\dagger}$ comment in the caption of Table \ref{Table Vacuum Case} for further details) gives us the values of $Q$ for which there is late-time isotropization; this yields $\frac{1}{2} <\left|Q\right|\leq1$. For the de Sitter case ($Q=\pm\frac{1}{2}$), the line $\mathcal{L}_{2}$ isotropizes as $t\to\infty$ due to the decaying exponential in $\sigma(t)$. We therefore conclude that on the interval that it is defined ($\left|Q\right|\leq1$), $\mathcal{L}_{2}$ is always isotropic for $Q=0$ (the static case) and isotropizes for $Q=\pm\frac{1}{2}$ (the de Sitter case). For the power-law case, isotropization only occurs when $\frac{1}{2} <\left|Q\right|\leq1$. As for the remaining thirteen fixed points, they all lie on the plane $\Sigma =0$ (and are hence all isotropic for all times) with the exception of $\mathcal{E}_{\pm}$, which are the only non-isotropic fixed points. Nonetheless, $\mathcal{E}_{\pm}$ also isotropize as $t\to\infty$.

\begin{figure}[h]  %
  \centering          
\includegraphics[width=0.7\textwidth]{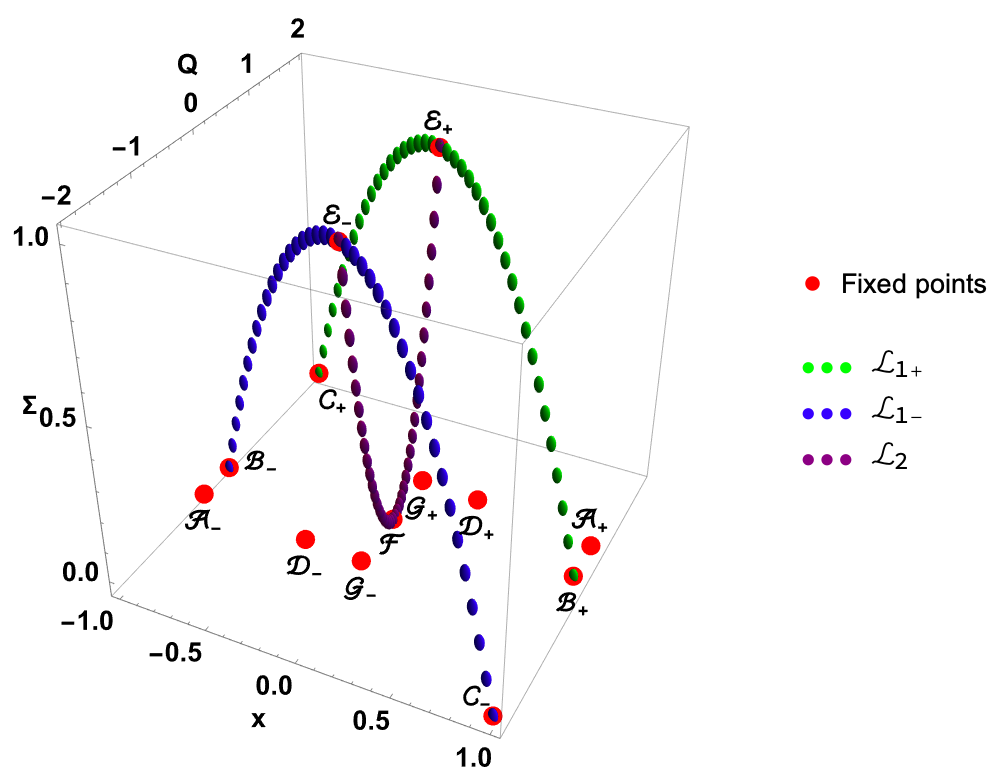} 
  \caption{The lines $\mathcal{L}_{1 \pm}$ and $\mathcal{L}_{2}$ and fixed points $\mathcal{A_{\pm}},\mathcal{B_{\pm}},\mathcal{C_{\pm}},\mathcal{D_{\pm}},\mathcal{E_{\pm}}, \mathcal{F} \text{ and } \mathcal{G_{\pm}}$ of the 3D vacuum system for the HS model with parameters $\{n,C_{1}\} = \{1,1\}$. Notice that the pair of lines $\mathcal{L}_{1\pm}$ (in green and blue respectively) meet the line $\mathcal{L}_{2}$ (in purple) on the $\Sigma=1$ plane. The line $\mathcal{L}_{1+}$ meets the line $\mathcal{L}_{2}$ at the fixed point $\mathcal{E}_{+}$, while the line $\mathcal{L}_{1-}$ meets the line $\mathcal{L}_{2}$ at the fixed point $\mathcal{E}_{-}$. The remaining eleven isotropic fixed points all lie on the $\Sigma=0$ plane (see Fig. \ref{Stability vacuum} in Appendix \ref{Appendix C} for a 2D slice of the system at $\Sigma=0$).}
  \label{DS Vacuum}
\end{figure}

Shown in Fig. \ref{DS Vacuum} is the 3D phase space of the system for the vacuum case with the lines $\mathcal{L}_{1\pm}$ and $\mathcal{L}_{2}$, alongside all the thirteen fixed points referred in Table \ref{Table Vacuum Case}.
As can be seen from Fig. \ref{DS Vacuum}, the lines $\mathcal{L}_{1\pm}$ are symmetric and both meet the other line $\mathcal{L}_{2}$ on the $\Sigma = 1$ plane. In particular, $\mathcal{L}_{1+}$ meets the line $\mathcal{L}_{2}$ at the fixed point $\mathcal{E}_{+}$ and $\mathcal{L}_{1-}$ meets the line $\mathcal{L}_{2}$ at the fixed point $\mathcal{E}_{-}$. Fig. \ref{DS Vacuum} also shows the eleven isotropic fixed points that all lie on the $\Sigma=0$ plane (see Fig. \ref{Stability vacuum} in Appendix \ref{Appendix C} for a 2D slice of the system at $\Sigma=0$). It is also worth noting that the line $\mathcal{L}_{1+}$ meets the fixed points $\mathcal{C}_{+}$ and $\mathcal{B}_{+}$ on the plane $\Sigma=0$; the line $\mathcal{L}_{1-}$ meets the fixed points $\mathcal{C}_{-}$ and $\mathcal{B}_{-}$ on the plane $\Sigma=0$. As for the line $\mathcal{L}_{2}$ it meets the $\Sigma=0$ plane once at $\mathcal{F}$. 

\begin{figure}[h]  %
  \centering          
  \includegraphics[width=0.95\textwidth]{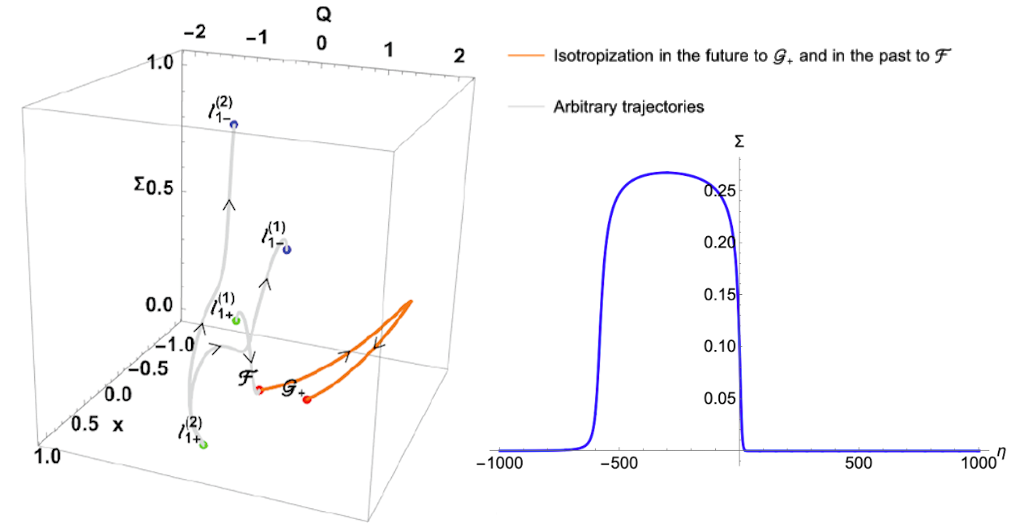}  
  \caption{\textbf{a.} (Left)
  Phase portrait of the 3D vacuum system for the HS model with parameters $\{n,C_{1}\} = \{1,1\}$ showing different trajectories. In grey, we have three arbitrary trajectories while in orange, we have one trajectory that starts (at $\eta=0$) at $(x,Q,\Sigma)=(-0.5,1.3,0.1)$ and  isotropizes in the future to the isotropic fixed point $\mathcal{G}_{+}$ and in the past to the isotropic fixed point $\mathcal{F}$. One of the arbitrary trajectories starts close to the fixed point $l^{(1)}_{1+}$, which lies on $\mathcal{L}_{1+}$, and evolves to $\mathcal{F}$. The remaining two arbitrary trajectories start close from the fixed point  $l^{(2)}_{1+}$, which also lies on $\mathcal{L}_{1+}$, and evolve to fixed points $l^{(1)}_{1-}$ and $l^{(2)}_{1-}$ (which both lie on $\mathcal{L}_{1-}$) respectively. \textbf{b.} (Right): Evolution of the shear $\Sigma$ against the phase-space time variable $\eta$ of the orange trajectory of Fig. \ref{Shear plot} (a). For both $\eta\rightarrow\pm\infty$, $\Sigma$ tends to $0$ showing isotropization in both the past and in the future.} 
  \label{Shear plot}
\end{figure}

A peculiar feature of $f(R)$ theories is that unlike standard GR, where isotropization can only occur in the future, the former theories allow for isotropization both in the future and in the past. This was shown for $f(R)= R^{n}$ in \cite{Leach:2006br}, where the authors found a trajectory that isotropizes in the past for $1<n<\frac{5}{4}$. Since in the appropriate limits, the HS $f(R)$ model can be approximated by $f(R)=R^{n}$, we also expect our system to exhibit past isotropization. A careful analysis of the phase-space trajectories reveals that the HS $f(R)$ model admits not only trajectories exhibiting past isotropization but also trajectories along which both past and future isotropization occur. Shown in Fig. \ref{Shear plot} (a) is the phase space showing different trajectories.
As can be seen from Fig. \ref{Shear plot} (a), the orange trajectory connects the two isotropic fixed points $\mathcal{G}_{+}$ and $\mathcal{F}$, isotropizing in the past towards $\mathcal{F}$ and in the future towards $\mathcal{G}_{+}$. Physically, this corresponds to a Universe that begins in an isotropic state (for e.g., at $\mathcal{F}$), evolves through a transient anisotropic phase, and eventually re-isotropizes as it expands, reaching $\mathcal{G}_{+}$. Such trajectories may therefore alleviate the anisotropy-related fine-tuning required for the onset of a subsequent inflationary phase. We found that trajectories that started close to the initial condition, $(x,Q,\Sigma)=(-0.5,1.3,0.1)$ could also isotropize in both directions, showing that the orange trajectory of Fig. \ref{Shear plot} (a) is not a special trajectory. The three arbitrary trajectories in Fig. \ref{Shear plot} (a) start close to the points $l^{(1)}_{1+}$ and $l^{(2)}_{1+}$ on $\mathcal{L}_{1+}$. One of them isotropizes to the point $\mathcal{F}$, while the other two evolve towards the fixed points $l^{(1)}_{1-}$ and $l^{(2)}_{1-}$ on $\mathcal{L}_{1-}$. Shown in Fig. \ref{Shear plot} (b) is the evolution of $\Sigma$ as a function of $\eta$ for the orange trajectory of Fig. \ref{Shear plot} (a) . As can be seen from Fig. \ref{Shear plot} (b), for both increasing and decreasing $\eta$, the shear $\Sigma$ tends to 0, implying isotropization in both the past and in the future.

%% file: 05-Matter.tex
\subsection{Matter Case}
\label{sec:Matter}
For the matter case, the upcoming analysis is further subdivided into the dust ($w=0$) and radiation ($w= \frac{1}{3}$) cases. The procedure remains the same as in the vacuum case, with the caveat that we can no longer reduce the 4D system of (\ref{Reg Compact DS}) into a 3D one (since $\Omega \ne 0$ here). Consequently, we have to solve the full 4D system, which will introduce not only lines of fixed points, but also 2D sheets of fixed points. 
\subsubsection{Dust}
For dust, we set $w=0$ in (\ref{Reg Compact DS}) to obtain our dynamical system, with the results shown in Table \ref{Table Dust}. 

\begin{table}[h]
	\centering
    \resizebox{\textwidth}{!}{%
  
	\begin{tabular}{|c|c|c|c|c|}
    
		\hline
		Point & $(\Sigma,x,v,Q,\Omega, y)$ & Stability & Scale factor, $a(t)$ & Shear, $\sigma(t)$\\
		
		\hline
		$\mathcal{S}_{1\pm}\,^{\dagger}$ & $\left(\Sigma, x,0,\pm 1-x,1-x^{2}-\Sigma,0 \right)$ & Unstable/Stable &
    $\begin{cases}

    & \text{$a_{0}(t-t_{0})^{\tfrac{(\mp1+x)^{2}}{2(\mp1+x)^{2}+\Sigma}}$},\\
    
    & \text{$a_{0}$ for $x = \pm1$}\\
    
    \end{cases}$
    
    & 
    $\begin{cases}
    
    & \text{$ \sigma_{0}\left[a_{0}(t-t_{0})^{\tfrac{(\mp1+x)^{2}}{2(\mp1+x)^{2}+\Sigma}}\right]^{\tfrac{\pm3-x}{\mp1+x}}$}, \\

    & \text{$0$ for $x=\pm1$ or $\Sigma = 0$}\\ 
    
    \end{cases}$
    \\
	    \hline

        $\mathcal{S}_{2}\,^{\dagger}$ & $\left( \Sigma,0,1-Q^{2},Q,Q^{2}-\Sigma,1-Q^{2} \right)$ & 
        
    \text{Saddle} &
    
    $\begin{cases}
    
    & \text{$a_{0}(t-t_{0})^{\frac{Q^{2}}{-1+3Q^{2}+\Sigma}}$}, \\
    & \text{$a_{0}{\rm e}^{H_{\rm dS}t}$ for $\tfrac{1}{2}\leq\left|Q\right| \leq \frac{1}{\sqrt{3}}$ },\\ 
    &\text{$a_{0}$ for $Q = 0$}
    
    \end{cases}$
    
    & 
    $\begin{cases}
    
    & \text{$\frac{\sigma_{0}}{{a_{0}^{3}}} {(t-t_{0})}^{-\frac{3Q^{2}}  {-1+3Q^{2}+\Sigma}}$}, \\
    
    & \text{$\frac{\sigma_{0}}{{a_{0}}^{3}}{\rm e}^{-3H_{\rm dS}t}$ for $\tfrac{1}{2}\leq\left|Q\right| \leq \frac{1}{\sqrt{3}}$ }, \\
     
    & \text{$0$ for $Q=0$ or $\Sigma=0$ }
    
    \end{cases}$
    \\

    \hline
		$\mathcal{L}_{3\pm}$ & $\left(0,x,0,\pm 1-x,1-x^{2},0 \right)$ & Unstable/Stable &
    $\begin{cases}
    
    & \text{$a_{0}\sqrt{t-t_{0}}$}, \\
    
    & \text{$a_{0}$ for $x = \pm1$}\\ 
    
    \end{cases}$
    
    & 0
    \\
    \hline

    $\mathcal{L}_{4}\,^{\dagger}$ & $\left(0,0,1-Q^{2},Q,Q^{2},1-Q^{2} \right)$ & 

    $\begin{cases}
        & \text{Saddle for $\left| Q \right| \lesssim0.33$},\\
        & \text{Unstable otherwise}
    \end{cases}$
    
    &
    $\begin{cases}
    
    & \text{$a_{0}(t-t_{0})^{\frac{Q^{2}}{-1+3Q^{2}}}$}, \\
    & \text{$a_{0}{\rm e}^{H_{\rm dS}t}$ for $Q=\pm\frac{1}{\sqrt{3}}$},\\
    & \text{$a_{0}$ for $Q = 0$}\\ 
    
    \end{cases}$
    
    & 0
    \\

    \hline
    $\mathcal{H}_{\pm}$ & $\left(0,\pm 1,0,\pm \frac{1}{2},0,-\frac{5}{4} \right)$ & Saddle & $a_{0}\sqrt{t-t_{0}}$ & 0\\

    \hline
    $\mathcal{I}_{\pm}$ & $\left(0,\pm 1,0,0,0,-\frac{5}{4} \right)$ & Unstable/Stable & $a_{0}$ & 0\\

    \hline

    $\mathcal{J}_{\pm}$ & $\left(0,\mp 1,0,\pm 2,0,0 \right)$ & Unstable/Saddle & $a_{0}\sqrt{t-t_{0}}$ & 0\\

    \hline

    $\mathcal{K_{\pm}}$ & $\left(0,\pm \frac{1}{3},0, \pm\frac{2}{3},\frac{8}{9},0 \right)$ & Unstable & $a_{0}\sqrt{t-t_{0}}$ & 0 \\

    \hline

    $\mathcal{M}_{\pm}$ & $\left( 0,\pm\frac{1}{3},\frac{8}{9},\pm\frac{2}{3},0,0\right)$ & Stable/Unstable & $a_{0} {\rm e}^{H_{\rm dS}t}$ & $0$  \\

    \hline

    $\mathcal{N_{\pm}}$ & $\left(0,0,0,\pm1,1,0 \right)$ & Unstable & $a_{0} \sqrt{t-t_{0}}$ & $0$  \\

    \hline

    $\mathcal{O_{\pm}}$ & $\left(0,0,\frac{1}{2},\pm \frac{1}{\sqrt{2}},0,\frac{1}{2} \right)$ & Unstable & $a_{0}(t-t_{0})$ & $0$  \\

    \hline

    $\mathcal{P_{\pm}}$ & $\left(0,0,1,\pm \frac{1}{\sqrt{2}},0,\frac{1}{2} \right)$ & Saddle/Unstable & $a_{0} {\rm e}^{H_{\rm dS}t}$ & $0$  \\

    \hline

    $\mathcal{Q_{\pm}}$ & $\left(0,-\frac{3}{\sqrt{11}},0,\pm1+ \frac{3}{\sqrt{11}},\frac{2}{11},0 \right)$ & Unstable & $a_{0} \sqrt{t-t_{0}}$ & $0$  \\

    \hline
    
    $\mathcal{R_{\pm}}$ & $\left(0,\frac{3}{\sqrt{11}},0,\pm1- \frac{3}{\sqrt{11}},\frac{2}{11},0 \right)$ & Unstable & $a_{0} \sqrt{t-t_{0}}$ & $0$  \\

    \hline

	    \end{tabular}}
		\caption{Fixed points for the HS model with the parameters $\{n,C_{1}\} = \{1,1\}$ for dust ($w = 0$). Notice that this system has three sheets of fixed points, $\mathcal{S}_{1\pm}$ and $\mathcal{S}_{2}$, and three lines of fixed points, $\mathcal{L}_{3\pm}$ and $\mathcal{L}_{4}$. Similar as before, where unspecified, the dynamical variables can take all the values on the interval over which they were defined in (\ref{Compact Variables Range}). $^{\dagger}$Here, for the sheets $\mathcal{S}_{1\pm}$, $\mathcal{S}_{2}$ and the line $\mathcal{L}_{4}$, the values of the dynamical variables are not their entire range. For the sheet $\mathcal{S}_{1\pm}$, the range for $x$ remains $|x|\leq 1$  with the added condition $\Sigma\in [0,1-x^{2}]$ as $\Omega = 1-x^{2}-\Sigma$ on the sheet; $\Omega$ and $\Sigma$ are both bounded on the interval $\left[0,1\right]$, which puts a restriction on the possible values $\Sigma$ can take. These conditions also imply that for the sheet $\mathcal{S}_{1}$, if $x=\pm1$, then $\Sigma=0$ by necessity for all conditions to hold. For the sheet $\mathcal{S}_{2}$, the range for $Q$ is $|Q| \leq 1$ with $\Sigma\in[0,Q^{2}]$ as $v = 1-Q^{2}$ and $\Omega=Q^{2}-\Sigma$ on the sheet; $v$ and $\Omega$ are both bounded on the interval $\left[0,1\right]$, which puts a restriction on the possible values $Q$ can take. These conditions also imply that for the sheet $\mathcal{S}_{2}$, if $Q=0$, then $\Sigma=0$ by necessity for all conditions to hold. For the line $\mathcal{L}_{4}$, we have $v=1-Q^{2}$ and $\Omega=Q^{2}$, which implies that $|Q| \leq 1$.}
		\label{Table Dust}
\end{table}
As can be seen from Table \ref{Table Dust}, the matter case for dust has three sheets of fixed points $\mathcal{S}_{1 \pm}$ and $\mathcal{S}_{2}$, three lines of fixed points, $\mathcal{L}_{3 \pm}$ and $\mathcal{L}_{4}$, and twenty fixed points, referred to as $\mathcal{H_{\pm}},\, \mathcal{I_{\pm}},\, \mathcal{J_{\pm}},\, \mathcal{K_{\pm}},\, \mathcal{M_{\pm}},\,  \mathcal{N_{\pm}},\, \mathcal{O_{\pm}},\, \mathcal{P_{\pm}},\, \mathcal{Q_{\pm}} \text{ and } \mathcal{R_{\pm}}$. The cosmic expansions for all the fixed points are still described, as in the vacuum case, by three scale factor solutions: a power-law expansion, a de Sitter expansion and a static universe. Similar as in the vacuum case, stability of the fixed points was determined by the eigenvalues of the Jacobian matrix of the system (see Appendix \ref{Appendix C} for further details).

All of the twenty fixed points, as well as the lines $\mathcal{L}_{3\pm}$ and $\mathcal{L}_{4}$, are isotropic for all times. Concerning the sheet $\mathcal{S}_{2}$, it is always isotropic for $Q=0$ (the static case) and isotropizes when $\tfrac{1}{2}\leq\left|Q\right| \leq \frac{1}{\sqrt{3}}$ (the de Sitter case). To find the values for which $\mathcal{S}_{2}$ can isotropize for the power-law case, we solve $-\frac{3Q^{2}}{-1+3Q^{2}+\Sigma}<0$ constraint to $|Q|\leq1$ and $0\leq\Sigma \leq Q^{2}$ (see $^{\dagger}$ under the caption of Table \ref{Table Dust} for details). Under these constraints, the inequality is solved for $0\leq\left|Q\right| \leq 1$, i.e., all values over which $\mathcal{S}_{2}$ is defined. However, as the range for which the power-law holds true is $Q\in\left[-1,1\right] \setminus \left(\left[-\frac{1}{\sqrt{3}},-\frac{1}{2}\right] \cup \left[\frac{1}{2},\frac{1}{\sqrt{3}}\right]  \right)$, the power-law case isotropizes on the latter interval only. A very similar calculation for $\mathcal{S}_{1\pm}$ shows that the power-law case will isotropize for $|x|\leq1$. However, this range of $x$ includes the static case ($x=\pm1$), which is always isotropic. Hence, the sheets $\mathcal{S}_{1\pm}$ will isotropize strictly when $\left|x\right|<1$  in the power-law case.

\begin{figure}[h]  %
  \centering          
  \includegraphics[width=0.6\textwidth]{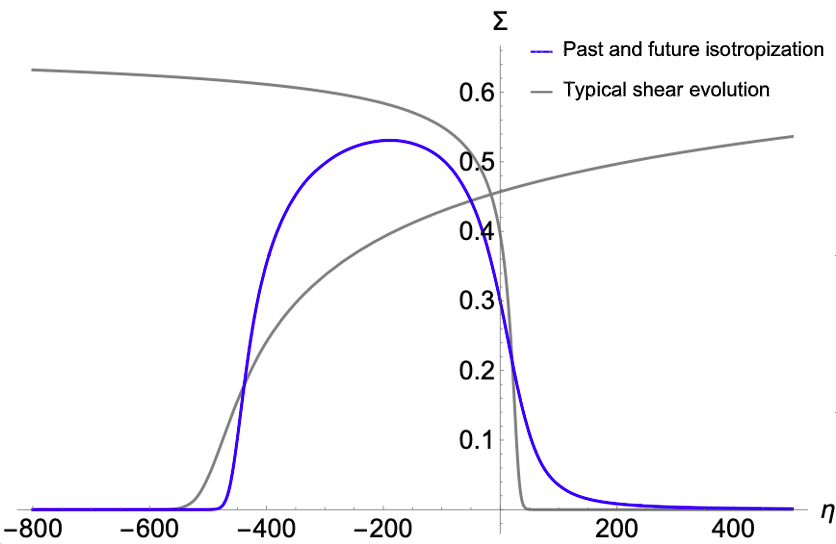}  
  \caption{ Evolution of the shear $\Sigma$ against the phase-space time variable $\eta$ for three trajectories in the dust case. In grey, we have the shear evolution for two arbitrary trajectories, while in blue we have the shear evolution for a trajectory that isotropizes both in the past and in the future. } 
  \label{Shear plot dust}
\end{figure}

Similar to the vacuum case, we were able to find a trajectory that isotropizes both in the past and in the future. Since the matter case is a 4D system, the explicit trajectory cannot be plotted, but we can nonetheless plot the shear evolution with the phase space time variable $\eta$. Shown in Fig. \ref{Shear plot dust} is the shear evolution for different trajectories in the dust case. As can be seen from Fig. \ref{Shear plot dust} the dust case admits trajectories that can isotropize both in the past and in the future. In particular, our analysis shows that the blue trajectory in Fig. \ref{Shear plot dust} starts (at $\eta=0$) at the initial condition $(x,v,Q,\Sigma)=(0.2,0.6,-0.85,0.3)$, and isotropizes in the future towards the line $\mathcal{L}_{4}$ at $Q=0$ and in the past to the isotropic fixed point $\mathcal{P}_{-}$. Similar as in the vacuum case, we found that trajectories that started out close to this initial condition could also isotropize in both directions.  

\FloatBarrier
\subsubsection{Radiation}
For radiation, we set $w=\frac{1}{3}$ in (\ref{Reg Compact DS}) to obtain our dynamical system, with the results shown in Table \ref{Table Radiation}.

\begin{table}[h]
	\centering
    \resizebox{\textwidth}{!}{%
  
	\begin{tabular}{|c|c|c|c|c|}
    
		\hline
		Point & $(\Sigma,x,v,Q,\Omega, y)$ & Stability & Scale factor, $a(t)$ & Shear, $\sigma(t)$\\
		
		\hline
		$\mathcal{S}_{1\pm}\,^{\dagger}$ & $\left(\Sigma, x,0,\pm 1-x,1-x^{2}-\Sigma,0 \right)$ & Unstable/Stable &
    $\begin{cases}

    & \text{$a_{0}(t-t_{0})^{\tfrac{(\mp1+x)^{2}}{2(\mp1+x)^{2}+\Sigma}}$},\\

    & \text{$a_{0}$ for $x = \pm1$}\\ 
    
    \end{cases}$
    
    & 
    $\begin{cases}

    & \text{$ \sigma_{0}\left[a_{0}(t-t_{0})^{\tfrac{(\mp1+x)^{2}}{2(\mp1+x)^{2}+\Sigma}}\right]^{\frac{\pm3-x}{\mp1+x}}$}, \\
    
    & \text{$0$ for $x=\pm1$ or $\Sigma = 0$}\\ 
    
    \end{cases}$
    \\
	    \hline

        $\mathcal{S}_{2}\,^{\dagger}$ & $\left( \Sigma,0,1-Q^{2},Q,Q^{2}-\Sigma,1-Q^{2} \right)$ & 
        
    \text{Saddle} &
    
    $\begin{cases}  
    & \text{$a_{0}(t-t_{0})^{\frac{Q^{2}}{-1+3Q^{2}+\Sigma}}$}, \\
    & \text{$a_{0}{\rm e}^{H_{\rm dS}t}$ for $\tfrac{1}{2}\leq\left|Q\right| \leq \frac{1}{\sqrt{3}}$ },\\ 
    &\text{$a_{0}$ for $Q = 0$}
    \end{cases}$
    
    & 
    $\begin{cases}
    
    & \text{$\frac{\sigma_{0}}{{a_{0}^{3}}} {(t-t_{0})}^{-\frac{3Q^{2}}  {-1+3Q^{2}+\Sigma}}$}, \\
    
    & \text{$\frac{\sigma_{0}}{{a_{0}}^{3}}{\rm e}^{-3H_{\rm dS}t}$ for $\tfrac{1}{2}\leq\left|Q\right| \leq \frac{1}{\sqrt{3}}$ }, \\
     
    & \text{$0$ for $Q=0$ or $\Sigma=0$ }
    \end{cases}$
    \\

    \hline
		$\mathcal{L}_{3\pm}$ & $\left(0,x,0,\pm 1-x,1-x^{2},0 \right)$ & Unstable/Stable &
    $\begin{cases}
    
    & \text{$a_{0}\sqrt{t-t_{0}}$}, \\
    
    & \text{$a_{0}$ for $x = \pm1$}\\ 
    
    \end{cases}$
    
    & 0
    \\
    \hline

    $\mathcal{L}_{4}\,^{\dagger}$ & $\left(0,0,1-Q^{2},Q,Q^{2},1-Q^{2} \right)$ & 

    $\begin{cases}
        & \text{Saddle for $\left| Q \right| \lesssim0.33$},\\
        & \text{Unstable otherwise}
    \end{cases}$
    
    &
    $\begin{cases}
    
    & \text{$a_{0}(t-t_{0})^{\frac{Q^{2}}{-1+3Q^{2}}}$}, \\
    & \text{$a_{0}{\rm e}^{H_{\rm dS}t}$ for $Q=\pm\tfrac{1}{\sqrt{3}}$},\\
    & \text{$a_{0}$ for $Q = 0$}\\ 
    
    \end{cases}$
    
    & 0
    \\

    \hline
    $\mathcal{L}_{5}$ & $\left(\Sigma,0,1,0,-\Sigma,1 \right)$ & Saddle & $a_{0}$ &
    
    $\begin{cases}
    
    & \text{Indeterminate}, \\
    & \text{$0$ for $\Sigma=0$}\\ 
    
    \end{cases}$
    \\
    
    \hline
    $\mathcal{H}_{\pm}$ & $\left(0,\pm 1,0,\pm \frac{1}{2},0,-\frac{5}{4} \right)$ & Saddle & $a_{0}\sqrt{t-t_{0}}$ & 0\\

    \hline
    $\mathcal{I}_{\pm}$ & $\left(0,\pm 1,0,0,0,-\frac{5}{4} \right)$ & Unstable/Stable & $a_{0}$ & 0\\

    \hline

    $\mathcal{J}_{\pm}$ & $\left(0,\mp 1,0,\pm 2,0,0 \right)$ & Unstable/Saddle & $a_{0}\sqrt{t-t_{0}}$ & 0\\

    \hline

    $\mathcal{K_{\pm}}$ & $\left(0,\pm \frac{1}{3},0, \pm\frac{2}{3},\frac{8}{9},0 \right)$ & Unstable & $a_{0} {\rm e}^{H_{\rm dS}t}$ & 0 \\

    \hline

    $\mathcal{M}_{\pm}$ & $\left( 0,\pm\frac{1}{3},\frac{8}{9},\pm\frac{2}{3},0,0\right)$ & Stable/Unstable & $a_{0} {\rm e}^{H_{\rm dS}t}$ & $0$  \\

    \hline

    $\mathcal{N_{\pm}}$ & $\left(0,\pm\frac{1}{3},0,\mp\frac{4}{3},\frac{8}{9},-\frac{16}{9} \right)$ & Unstable & $
    a_{0}\sqrt{{t-t_{0}}}$ & $0$  \\

    \hline

    $\mathcal{O_{\pm}}$ & $\left(0,0,\frac{1}{2},\pm \frac{1}{\sqrt{2}},0,\frac{1}{2} \right)$ & Unstable & $a_{0}(t-t_{0})$ & $0$  \\

    \hline

    $\mathcal{P_{\pm}}$ & $\left(0,0,1,\pm \frac{1}{\sqrt{2}},0,\frac{1}{2} \right)$ & Saddle/Unstable & $a_{0}{\rm e}^{H_{\rm dS}t}$ & $0$  \\

    \hline

	    \end{tabular}}
		\caption{Fixed points for the HS model with the parameters $\{n,C_{1}\} = \{1,1\}$ for radiation ($w = \frac{1}{3}$). Notice that this system has three sheets of fixed points, $\mathcal{S}_{1\pm}$ and $\mathcal{S}_{2}$, and four lines of fixed points, $\mathcal{L}_{3\pm}$, $\mathcal{L}_{4}$ and $\mathcal{L}_{5}$. $^{\dagger}$Here, for the sheets $\mathcal{S}_{1\pm}$, $\mathcal{S}_{2}$ and the line $\mathcal{L}_{4}$, the values of the dynamical variables are not their entire range, as was the case in the dust case. Since these are the same sheets and line as in the dust case, the restrictions of the dynamical variables remain the same for each of them. See $^{\dagger}$ under the caption of Table \ref{Table Dust} for details.}
		\label{Table Radiation}
\end{table}
As can be seen from Table \ref{Table Radiation}, the matter case for radiation has three sheets of fixed points $\mathcal{S}_{1 \pm}$ and $\mathcal{S}_{2}$, four lines of fixed points $\mathcal{L}_{3 \pm}$,  $\mathcal{L}_{4}$ and $\mathcal{L}_{5}$  and sixteen fixed points $\mathcal{H_{\pm}},\,\mathcal{I_{\pm}},\,\mathcal{J_{\pm}},\,\mathcal{K_{\pm}},\,\mathcal{M_{\pm}},\, \mathcal{N_{\pm}},\, \mathcal{O_{\pm}} \text{ and } \mathcal{P_{\pm}}$. The cosmic expansions for all fixed points are still described by the same three scale factor solutions we had in the dust case. The sheets $\mathcal{S}_{1 \pm}$ and $\mathcal{S}_{2}$ and the lines $\mathcal{L}_{3 \pm}$ and $\mathcal{L}_{4}$ repeat again from the dust case and have the same exact behavior as before. A novel line that did not appear in the dust case is $\mathcal{L}_{5}$. The line $\mathcal{L}_{5}$ is peculiar in the sense that its shear cannot be uniquely determined for all $\Sigma\ne0$. To see this explicitly, notice that for the line $\mathcal{L}_{5}$, we have $x=Q=0$, which results in an indeterminate form in (\ref{Gamma Definition}). Nonetheless, a careful analysis shows that this indeterminacy is merely a reflection of the fact that $\mathcal{L}_{5}$ is actually an unphysical line, save for when $\Sigma=0$. The latter fact can be seen by noticing that $\mathcal{L}_{5}$ is parametrized by $\Omega=-\Sigma$. However, both $\Omega$ and $\Sigma$ are bounded on the interval $\left[0,1\right]$, which means that the only value for which $\Omega=-\Sigma$ can hold true is $\Sigma=0$. Hence, the only physical part of $\mathcal{L}_{5}$ is the point $l_{5}$, with coordinates $(\Sigma,x,v,Q,\Omega, y)=\left(0,0,1,0,0,1\right)$. 

All lines and all of the fixed points are isotropic for all times. In the case of the sheets $\mathcal{S}_{1 \pm}$ and  $\mathcal{S}_{2}$, they repeat from the dust case and showcase the same behavior as before: the power-law case for $\mathcal{S}_{1\pm}$ isotropizes when $\left|x\right|<1$; $\mathcal{S}_{2}$ isotropizes for the de Sitter case ($\tfrac{1}{2}\leq\left|Q\right| \leq \frac{1}{\sqrt{3}}$) and for the power-law case when $Q\in\left[-1,1\right] \setminus \left(\left[-\frac{1}{\sqrt{3}},-\frac{1}{2}\right] \cup \left[\frac{1}{2},\frac{1}{\sqrt{3}}\right]\right)$.

\begin{figure}[h]  %
  \centering          
  \includegraphics[width=0.6\textwidth]{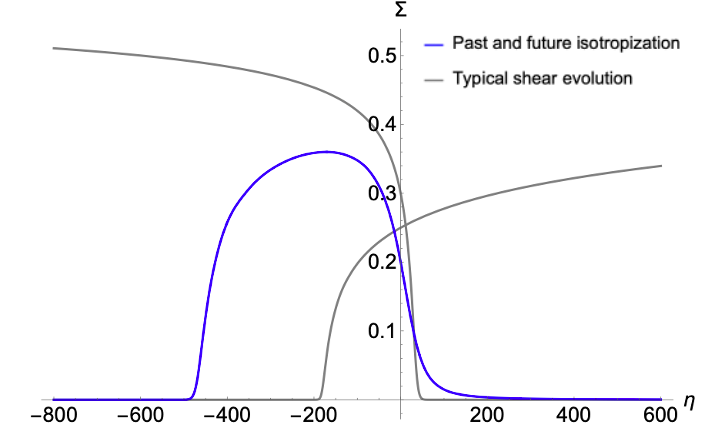}  
  \caption{ Evolution of the shear $\Sigma$ against the phase-space time variable $\eta$ for three trajectories in the radiation case. In grey, we have the shear evolution for two arbitrary trajectories, while in blue we have the shear evolution for a trajectory that isotropizes both in the past and in the future. } 
  \label{Shear plot rad}
\end{figure}

Shown in Fig. \ref{Shear plot rad} is the shear evolution for different trajectories in the radiation case. As can be seen from Fig. \ref{Shear plot rad} the radiation case also admits trajectories that can isotropize both in the past and in the future. In particular, our analysis shows that the blue trajectory in Fig. \ref{Shear plot rad} starts (at $\eta=0$) at the initial condition $(x,v,Q,\Sigma)=(0.25,0.6,-0.9,0.2)$, and isotropizes in the future towards the fixed point $l_{5}$ on $\mathcal{L}_{5}$ and in the past to the isotropic fixed point $\mathcal{P}_{-}$. Similar to the vacuum and dust cases, we found that trajectories that started out close to this initial condition could also isotropize in both directions.

\FloatBarrier

%% file: 06-Conclusion.tex
\section{Conclusions}
In this work we have presented a compact dynamical system analysis 
for Bianchi I spacetime embedded in $f(R)$ Hu-Sawicki gravity, specifically for the parameters $\{n,C_1\}=\{1,1\}$. We began our analysis by first providing a brief review of  $f(R)$ gravity followed by highlighting the field equations that hold in Bianchi I. We also included a succinct review of Hu-Sawicki gravity. We then defined six compact normalized dynamical variables that allowed us to derive a set of first-order autonomous ordinary differential equations. This system was analyzed in both the vacuum and matter cases to determine the fixed points, cosmic expansion and shear. First, for the vacuum case, we found three lines of fixed points and thirteen fixed points that had a cosmological history that evolved according to a power-law, de Sitter or static solution. Then, the matter case was further subdivided into the dust and radiation cases. A novel aspect of the matter cases is the emergence of 2D sheets of fixed points. Indeed, for the dust case, we found three sheets, three lines and twenty fixed points; for the radiation case three sheets, four lines and sixteen fixed points.

Subsequently, we found that in the vacuum case, all fixed points were either isotropic or could isotropize with time. We were also able to find trajectories that isotropize both in the past and in the future, a feature that is not present in Bianchi I when studied within the standard General Relativity framework.
For the matter case, we found that the only non-isotropic fixed points turned out to be the sheets $\mathcal{S}_{1\pm}$ and $\mathcal{S}_{2}$ and they could both isotropize for both cases studied (dust and radiation). Our key result is that for the Hu--Sawicki model with $n=C_1=1$, the compact Bianchi~I dynamical system admits trajectories that isotropize toward both the past and the future. In particular, modified-gravity dynamics can possess a past-isotropic sector which is absent in the corresponding Bianchi~I dynamics of GR. This provides a possible mechanism which may alleviate the anisotropy-related fine tuning required for the onset of a subsequent
inflationary phase, although it does not by itself specify the inflationary mechanism or establish a complete cosmological history from the past-isotropic regime to late-time acceleration.

A promising direction for future work would be to explore whether a trajectory that isotropizes both in the past and in the future naturally emerges in any arbitrary $f(R)$ theory. Since it is possible to study $f(R)$ cosmologies in a model-independent way by combining the dynamical systems approach with cosmographic parameters \cite{Capozziello:2008qc,Chakraborty:2021jku}, it should, in principle, be possible to deduce whether this feature is generic in $f(R)$ theories.

%% file: 07-Acknowledgements.tex
\section*{Acknowledgements}
AN acknowledges partial financial support from the University of Cape Town (UCT) through the UCT International and Refugee Scholarship.
AdlCD acknowledges support from BG20/00236 action (MCINU, Spain), NRF Grant CSUR23042798041, CSIC Grant COOPB23096, Project SA097P24 funded by Junta
de Castilla y Le\'on (Spain) and Grants PID2024-158938NB-I00 and PID2021-122938NB-I00 funded by MCIN/AEI/10.13039/501100011033 and by {\it ERDF A way of making Europe}.

%% file: 08-Appendix.tex
\begin{appendix}
\section{Derivation of $a(t)$}
\label{Appendix A}

To derive an expression for $a(t)$, we start by substituting (\ref{Friedmann}) into (\ref{Raychaudauri}). This gives, after simplification, 
\begin{equation}
    2\dot{H}+\frac{1}{f'}\left(\rho-3Hf''\dot{R}\right)+2\sigma^{2} = -\frac{1}{f'}\left[w\rho+\dot{R}^{2}f'''+\left(2H\dot{R}+\ddot{R}\right)f''\right].
    \label{Continuity Eq}
\end{equation}
The goal is to recast every single term in terms of the normalized variables defined in (\ref{compact variables}) and $H$ so that the resulting equation can be easily integrated for $a(t)$. To this end, we notice that (\ref{Continuity Eq}) can be re-written as
\begin{equation}
    2\dot{H}+\frac{\rho}{f'}(1+w)-\frac{Hf''\dot{R}}{f'}+2\sigma^{2}=-\frac{1}{f'}\left(\dot{R}^{2}f'''+\ddot{R}f''\right).
    \label{Continuty Eq2}
\end{equation}
To further simplify the RHS of (\ref{Continuty Eq2}), we need to make use of the trace equation, (\ref{Trace}). Using (\ref{Trace}), one has that 
\begin{equation}
    R = \frac{1}{f'}\left[3p-\rho+2f+3 \left(f''\ddot{R}+3H\dot{R} f''+f'''\dot{R}^{2} \right) \right].
    \label{Trace eq explicit}
\end{equation}
We simplify (\ref{Trace eq explicit}) using the normalized variables to find that 
\begin{equation}
    \frac{1}{f'}\left(\dot{R}^{2}f'''+\ddot{R}f''\right) = \frac{H^{2}}{Q^{2}}\left[2v-\Omega(3w-1)-4y-6xQ\right],
    \label{LHS Continuity eq}
\end{equation}
which corresponds to the RHS of (\ref{Continuty Eq2}) up to a negative sign. The LHS of (\ref{Continuty Eq2}) can similarly be written in terms of the normalized variables. Combining the latter result with (\ref{LHS Continuity eq}), (\ref{Continuty Eq2}) becomes
\begin{equation}
    2\dot{H}+\frac{3\Omega (1+w)}{Q^{2}}H^{2}-\frac{2x}{Q}H^{2}+\frac{6\Sigma}{Q^{2}}H^{2} = -\frac{H^{2}}{Q^{2}}\left[2v-\Omega(3w-1)-4y-6xQ\right].
    \label{Continuty Eq3}
\end{equation}
Upon combining all $H^{2}$ terms in (\ref{Continuty Eq3}) and simplifying, we obtain
\begin{equation}
    2\dot{H} + \alpha H^{2}=0,
    \label{H main eq}
\end{equation}
where $\alpha$ is defined as in (\ref{Alpha Definition}).
(\ref{H main eq}) can be easily integrated to find that 
\begin{equation}
    H =\frac{2}{\alpha(t-t_{0})}.
    \label{integrated H}
\end{equation}
Since $H=\frac{\dot{a}}{a}$, (\ref{integrated H}) can be further integrated to find that  
\begin{equation}
    a(t)=a_{0}(t-t_{0})^{\frac{2}{\alpha}},
    \label{scale factor def}
\end{equation}
as previously claimed. 

\section{Derivation of $\sigma(t)$}
\label{Appendix B}
To derive $\sigma(t)$, we first recast (\ref{Gauss-Codazzi}), in terms of the normalised variables and $H$, with the goal of obtaining an equation that can be easily integrated. As such, (\ref{Gauss-Codazzi}) becomes 
\begin{equation}
    \frac{\dot{\sigma}}{\sigma}=-\left(3+\frac{2x}{Q}\right)H.
    \label{GC variables}
\end{equation}
Again, since $H= \frac{\dot{a}}{a}$, this can be substituted in (\ref{GC variables}) and integrated to find that 
\begin{equation}
    \sigma(t) = \sigma_{0}a(t)^{-\gamma},
    \label{sigma def}
\end{equation}
where $\gamma$ is defined as in (\ref{Gamma Definition}). Since we have the analytical form for the scale factor, (\ref{scale factor def}), (\ref{sigma def}) becomes 
\begin{equation}
    \sigma(t) = \sigma_{0}{a_{0}^{-\gamma}(t-t_{0})^{-\frac{2\gamma}{\alpha}}},
    \label{Shear App}
\end{equation}
as previously claimed. Notice that for $Q=0$ and $\alpha=0$, \eqref{Shear App} becomes ill-defined and one must use the analytical expression that the scale factor takes (Einstein static for $Q=0$ and de Sitter for $\alpha=0$) and substitute in \eqref{sigma def} to obtain the shear at such points. 

\section{Stability}
\label{Appendix C}
As previously stated, the stability analysis was performed by first determining the eigenvalues of the Jacobian matrix for each system (vacuum, dust and radiation). Nonetheless, using the eigenvalues of the Jacobian alone was not sufficient to determine the stability of all the fixed points as some of them had vanishing eigenvalues, which meant that in these cases, this method failed. For such cases, other methods had to be employed.

An alternative method was to
expand the dynamical variables around a fixed point, for e.g., in the vacuum case: $x = x_{0} + \delta x$, $Q = Q_{0} + \delta Q$ and $\Sigma = \Sigma_{0} + \delta\Sigma$, where the  subscript $0$ denotes the coordinates of a fixed point and $\delta$ a small perturbation around it. Substituting into the evolution equations (\ref{Reg Compact DS}) and keeping terms up to first order yielded a closed system for $\delta x$, $\delta Q$ and $\delta \Sigma$, which could be solved exactly. This method allowed us to determine the stability of some fixed points not captured by the Jacobian analysis. However, it still failed when the first-order perturbations were found to be constant. Extending the expansion to second-order required a numerical treatment, but we found that it suffered the same limitation.

\begin{figure}[H]
    \centering
    \includegraphics[width=0.5\textwidth]{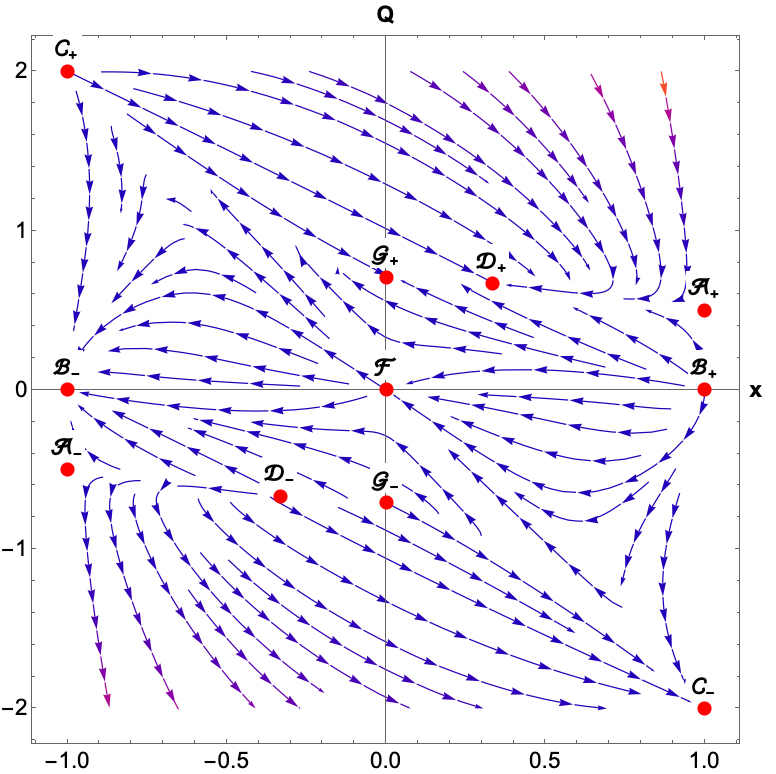} 
    \caption{Phase portrait of the dynamical system for the vacuum case ($\Omega=0$) sliced at $\Sigma =0$. Shown on the plot are the eleven isotropic fixed points $\mathcal{A}_{\pm}$, $\mathcal{B}_{\pm}$, $\mathcal{C}_{\pm}$, $\mathcal{D}_{\pm}$, $\mathcal{F}$ and $\mathcal{G}_{\pm}$. The stability of each of these eleven fixed points can be easily determined by virtue of the trajectories.}
    \label{Stability vacuum}
\end{figure}

The most effective method  for determining stability was to examine trajectories in the vicinity of each fixed point. Convergent trajectories indicated stability, divergent ones instability, and mixed behaviour identified saddle points. This method, of course, relied on the fact whether the trajectories could be plotted, which is feasible in the vacuum 3D case, but not directly in the matter 4D case. However, since $\Sigma =0$ and $v=0$ define invariant sub-manifolds, the system could be reduced to 2D or 3D slices where trajectories remained confined, enabling stability analysis. For example, in the vacuum case, slicing the 3D system at $\Sigma=0$ yielded a 2D phase portrait, from which fixed-point stability analysis can be read off. Shown in Fig. \ref{Stability vacuum} is the latter plot. 
Fig. \ref{Stability vacuum} shows that $\mathcal{A}_{\pm}$ are saddle points, while $\mathcal{B}_{\pm}$, $\mathcal{C}_{\pm}$, and $\mathcal{D}_{\pm}$ exhibit mixed stability: $\mathcal{B}_{+}$ ($\mathcal{B}_{-}$) is unstable (stable), $\mathcal{C}_{+}$ ($\mathcal{C}_{-}$) is unstable (stable), and $\mathcal{D}_{+}$ ($\mathcal{D}_{-}$) is stable (unstable). Point $\mathcal{F}$ is a saddle, and $\mathcal{G}_{\pm}$ follow the pattern $\mathcal{G}_{+}$ stable and $\mathcal{G}_{-}$ unstable.

\begin{figure}[H]
    \centering
    \begin{minipage}{5.9cm} 
        \centering
        \includegraphics[width=\linewidth]{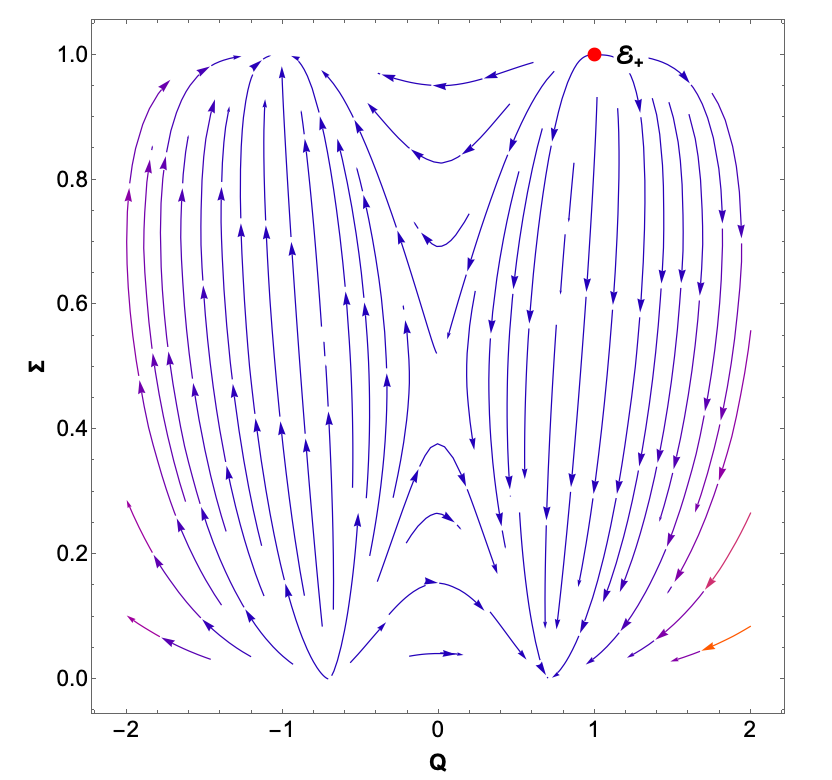}

        \label{fig:plot1}
    \end{minipage}\hfill 
    \begin{minipage}{5.9cm}
        \centering
        \includegraphics[width=\linewidth]{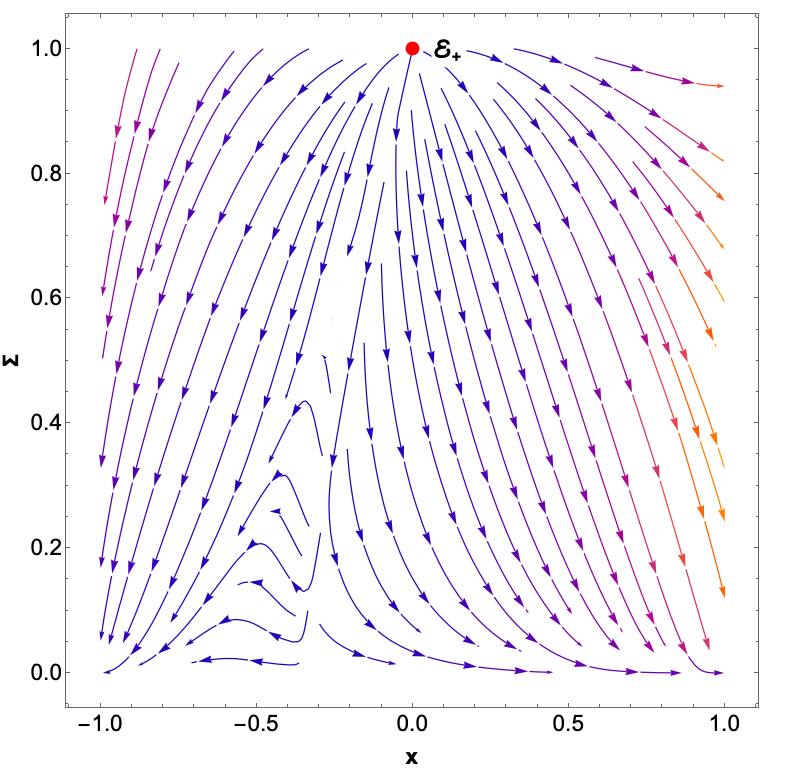}
        
        \label{fig:plot2}
    \end{minipage}\hfill
    \begin{minipage}{5.9cm}
        \centering
        \includegraphics[width=\linewidth]{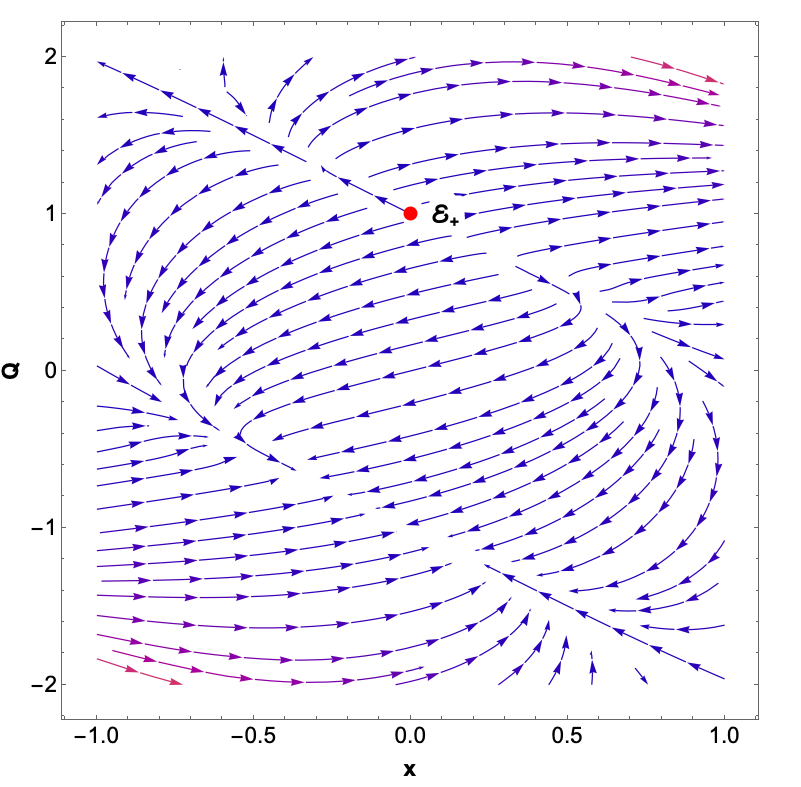}
    
        \label{fig:plot3}
    \end{minipage}
    \caption{The phase space sliced at the point $\mathcal{E}_{+}$ along the three axes. Shown (from left to right) are the $\Sigma-Q$, $\Sigma-x$ and $Q-x$ planes at that point. As can be seen from all three plots, the point $\mathcal{E}_{+}$ is unstable in all three directions.}
    \label{E_pm}
\end{figure}

\begin{figure}[H]

  \centering
  \begin{minipage}[t]{0.51\textwidth}
    \centering
    \includegraphics[width=\linewidth]{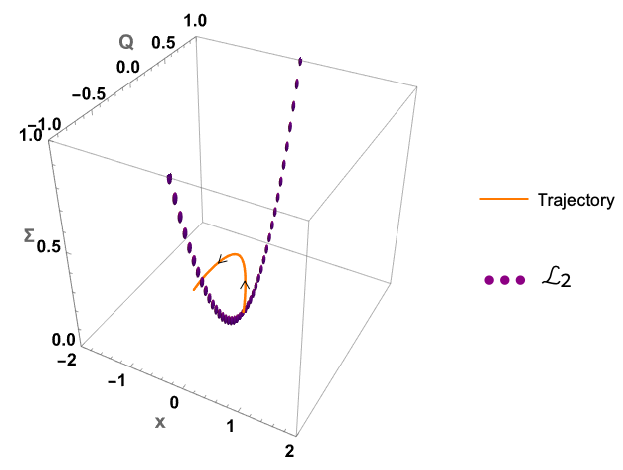}
  \end{minipage}%
  \quad
  \begin{minipage}[t]{0.37\textwidth}
    \centering
    \includegraphics[width=\linewidth]{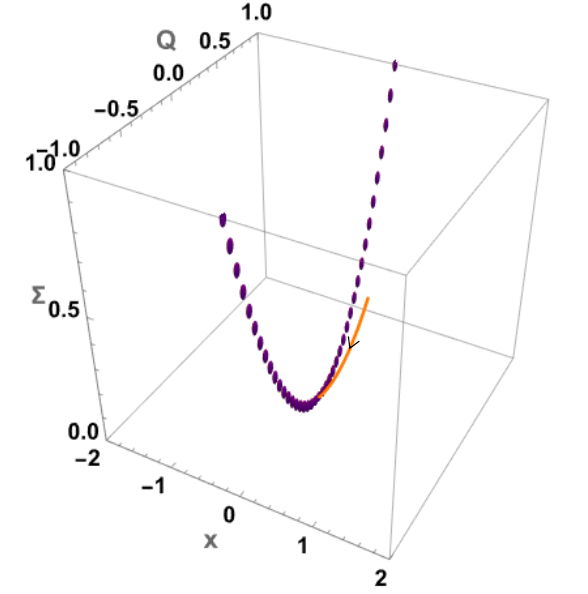}
  \end{minipage}
  \caption{Small perturbation around the line $\mathcal{L}_{2}$ (in purple) for the vacuum case. In the left plot, we perturb about $Q=0$ and find that the trajectory (in orange) diverges while in the right plot, we perturb about $Q=0.1$ and find that the trajectory converges. This analysis is repeated for many more such points to conclude that $\mathcal{L}_{2}$ is a saddle for $0\leq Q \lesssim 0.37$ and unstable otherwise.}
  \label{L_2 trajec}
\end{figure}

Notice that the points $\mathcal{E}_{\pm}$ do not appear in Fig. \ref{Stability vacuum} as they are the only non-isotropic fixed point of the vacuum case (i.e. $\Sigma \ne 0$ for $\mathcal{E}_{\pm}$). In this case, the 3D system was sliced in all three directions of the phase space at the points $\mathcal{E}_{\pm}$ to determine their stability. Shown in Fig. \ref{E_pm} are the $\Sigma-Q$, $\Sigma-x$ and $Q-x$ planes sliced at $\mathcal{E}_{+}$, which clearly demonstrates that $\mathcal{E}_{+}$ is unstable. 
Combining all the three methods outlined here, we were able to determine the stability of all the fixed points. For the three lines $\mathcal{L}_{1\pm}$ and $\mathcal{L}_{2}$, we looked at trajectories for small perturbation close to the lines. Shown in Fig. \ref{L_2 trajec} is an example of such perturbations around the line $\mathcal{L}_{2}$.
We repeated the analysis shown in Fig. \ref{L_2 trajec} for many more such perturbations around the line for all values $Q$ could take (here, $|Q|\leq 1$). Our analysis showed that $\mathcal{L}_{2}$ is a saddle for $0\leq Q \lesssim 0.37$ and unstable otherwise.

\begin{figure}[H]
  \centering
  
  \begin{minipage}[t]{0.57\textwidth}
    \centering
    \includegraphics[width=\linewidth]{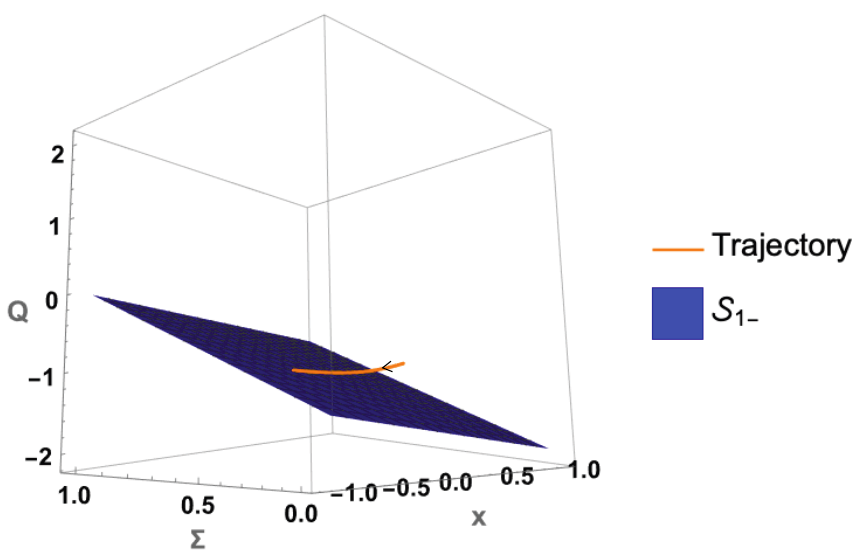}
    
    \label{fig:plot1}
  \end{minipage}
  \hfill
  \begin{minipage}[t]{0.37\textwidth}
    \centering
    \includegraphics[width=\linewidth]{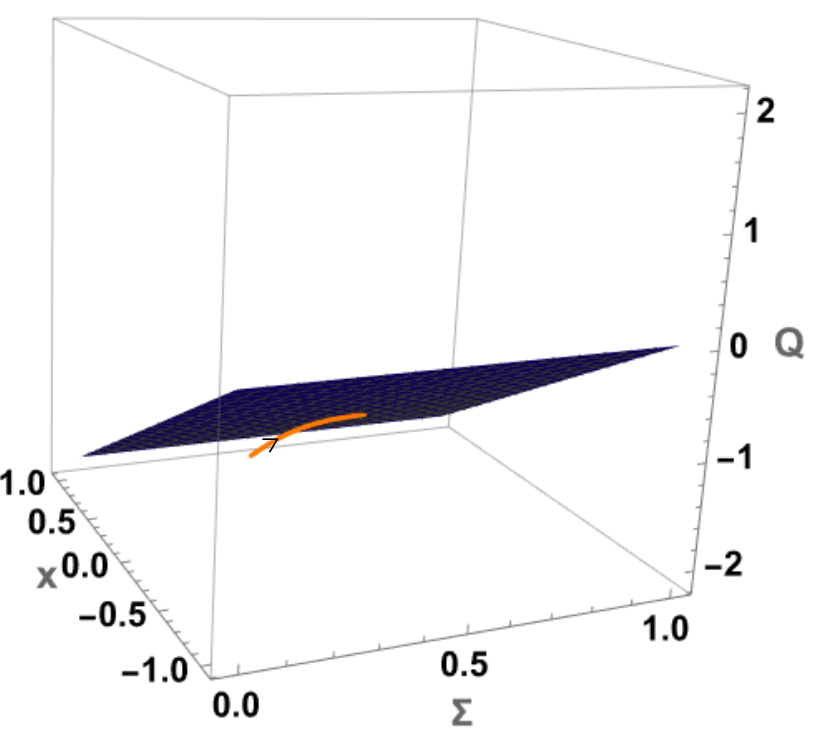}
    
  \end{minipage}
  \caption{Small perturbation around the sheet $\mathcal{S}_{1-}$ (in blue) of the matter case for dust sliced at the invariant sub-manifold $v=0$. In the left plot, we perturb `above' the sheet and find that the trajectory (in orange) converges and in the right plot, we perturb `below' the sheet and find that the trajectory also converges. This analysis is repeated for many more such points to conclude that $\mathcal{S}_{1-}$ is stable}
\label{Sheets trajectory}
\end{figure}

This analysis was similarly performed for the matter case (dust and radiation). We again made use of the fact that $\Sigma=0$ and $v=0$ are invariant sub-manifolds to determine stability whenever possible. When this was not feasible (for e.g the sheet $\mathcal{S}_{2}$, where $\Sigma\ne0$ and $v\ne0$), we simply plotted how the relevant dynamical variables evolved with the phase-space time variable $\eta$ when a small perturbation was applied. The only new aspect of the matter case involved 2D sheets of fixed points whose stability were determined analogously to the lines, by perturbing around the sheets and observing whether the resulting trajectories converged or diverged. Shown in Fig. \ref{Sheets trajectory} is the sheet $\mathcal{S}_{1-}$ in the dust case sliced at the invariant sub-manifold $v=0$.

We again repeated the analysis shown in Fig. \ref{Sheets trajectory} for many more such perturbations around the sheet. Our analysis showed that $\mathcal{S}_{1-}$ is stable.
\end{appendix}